\documentclass[a4paper,11pt]{article}
\UseRawInputEncoding
\usepackage{jheppub}
\usepackage{float}
\usepackage{amsmath,amssymb}
\usepackage{bm}
\usepackage{slashed}
\usepackage{epsfig}
\usepackage{graphicx}
\usepackage{hyperref}
\usepackage{xcolor}
\usepackage{braket}
\usepackage{subfigure}
\usepackage{soul}
\usepackage{mathtools}

\hypersetup{
    colorlinks=true,
    linkcolor=blue,
    filecolor=magenta,      
    urlcolor=blue,
    citecolor=blue
}

\begin{document}
\title{Partonic distribution functions and amplitudes using tensor network methods}

\author[a,b,c,d]{Zhong-Bo Kang,}
\emailAdd{zkang@physics.ucla.edu}
\affiliation[a]{Department of Physics and Astronomy, University of California,
Los Angeles, CA 90095, USA}
\affiliation[b]{Mani L. Bhaumik Institute for Theoretical Physics, University of California, Los Angeles, CA 90095, USA}
\affiliation[c]{Center for Quantum Science and Engineering, University of California, Los Angeles, CA 90095, USA}
\affiliation[d]{Center for Frontiers in Nuclear Science, Stony Brook University,
Stony Brook, NY 11794, USA}
\author[a,b]{Noah Moran,}
\emailAdd{chucknoah@g.ucla.edu}

\author[a]{Peter Nguyen,}
\emailAdd{peternguyen0128@g.ucla.edu}

\author[e]{and Wenyang Qian}
\emailAdd{qian.wenyang@usc.es}
\affiliation[e]{Instituto Galego de F\'isica de Altas Enerx\'ias IGFAE, Universidade de Santiago de Compostela,
E-15782 Galicia-Spain}

\abstract{
Calculations of the parton distribution function (PDF) and distribution amplitude (DA) are highly relevant to core experimental programs as they provide non-perturbative inputs to inclusive and exclusive processes, respectively. Direct computation of the PDFs and DAs remains challenging because they are non-perturbative quantities defined as light-cone correlators of quark and gluon fields, and are therefore inherently time-dependent. In this work, we use a uniform quantum strategy based on tensor network simulation techniques to directly extract these hadronic quantities from first principles using the matrix product state of the hadrons in the same setup. We present exemplary numerical calculations with the Nambu--Jona-Lasinio model in 1+1 dimensions and compare with available exact diagonalization and quantum circuit simulation results. Using tensor networks, we evaluate the PDF and DA at various strong couplings in the large-qubit limit, which is consistent with expectations at perturbative and non-relativistic limits.
}

\maketitle

\section{Introduction}
The detailed understanding of the internal structure of nucleons in terms of fundamental quarks and gluons remains a cornerstone of research in high-energy nuclear and particle physics~\cite{AbdulKhalek:2021gbh,Accardi:2012qut}. These structures are typically described using parton distribution functions (PDFs) and their extensions, such as transverse momentum-dependent parton distribution functions (TMDs), generalized parton distributions (GPDs), and parton distribution amplitudes (DAs). These functions are central to connecting theoretical predictions with experimental measurements in high-energy scattering processes. For example, PDFs play a critical role in predicting cross sections for processes such as single inclusive jet production in proton-proton collisions at colliders like the Large Hadron Collider (LHC). On the other hand, parton distribution amplitudes are key quantities in exclusive processes such as $\gamma^*\gamma \to \pi^0$.

Over the years, two principal approaches have been developed to determine non-perturbative parton distributions, such as PDFs and DAs, providing critical tools for exploring the internal structure of nucleons. The first approach is global QCD analysis, which exploits the QCD factorization framework~\cite{Collins:1989gx}. Within this framework, cross sections of physical processes are expressed as a convolution of perturbative short-distance coefficients and non-perturbative parton distributions. By analyzing diverse experimental cross-section data sets in conjunction with state-of-the-art perturbative calculations, PDFs or DAs can be extracted. This method has proven highly effective in incorporating experimental inputs to achieve precise determinations of parton distributions.

The second approach is based on lattice QCD, a method that allows for the first-principles computation of the QCD path integral in a discretized finite-volume Euclidean space-time. Direct lattice calculations of PDFs and DAs have traditionally been challenging because these quantities are defined as light-cone correlators, which are inherently time-dependent and thus difficult to compute in a Euclidean framework. However, significant progress has been made in recent years with the development of methods to compute quasi-PDFs~\cite{Ji:2013dva,Ji:2020ect} and pseudo-PDFs~\cite{Radyushkin:2017cyf}, which are related to the standard light-cone PDFs through rigorous theoretical frameworks. These breakthroughs have enabled the indirect extraction of PDFs from lattice computations, representing a major advance in the field~\cite{Lin:2017snn}. See other approaches such as the hadronic tensors~\cite{Liu:1993cv}, axuiliary quarks~\cite{Detmold:2005gg}, current-current correlators~\cite{Braun:2007wv,Ma:2017pxb}, operator product expansion~\cite{Chambers:2017dov}, and light-front Hamiltonian~\cite{Brodsky:2006uqa, 1stBLFQ}, as well as Refs~\cite{Ji:2020ect,Cichy:2018mum,Constantinou:2020pek} for a review on the progress of recent developments.

Quantum computing technology has undergone rapid development and may provide an alternative path toward calculating the structure of hadrons from first principles~\cite{Bauer:2022hpo, DiMeglio:2023nsa}. The exponential Hilbert space and quantum simulation algorithms provide a natural and efficient theoretical framework to simulate quantum field theory~\cite{Jordan:2011ci, Jordan:2014tma, Jordan:2017lea}. Studying real-time evolution, such as the case for the PDF and DA, is particularly suited to quantum simulation~\cite{Lamm:2019uyc}. The idea is to encode the fermionic and gluonic degrees of freedom into spin Hamiltonians and simulate the real-time matrix elements for evaluating these distribution functions and amplitudes directly on the quantum circuit.
Recently, exploratory studies using quantum computing for parton structures are booming: from low-dimensional (1+1)-dimensional field theories~\cite{Lamm:2019uyc,Perez-Salinas:2020nem,LiTianyin:2021kcs, Li:2022lyt, LiTianyin:2024nod, Grieninger:2024cdl,Grieninger:2024axp}, worldline formulation~\cite{Mueller:2019qqj}, spacetime Wilson loop~\cite{Echevarria:2020wct}, to effective light-front Hamiltonian models~\cite{Kreshchuk:2020dla,Kreshchuk:2020aiq,Qian:2021jxp}, these studies already exhibit interesting physics and potential. However, most quantum simulation for parton structure is still restricted in performing systematic calculation and limited in extending to the continuum, largely due to today's noisy intermediate-scale quantum infrastructure~\cite{Preskill:2018jim}.

In this work, we aim to leverage tensor network (TN) methods in computing hadron structures. Traditionally used in condensed matter physics, TN have become an increasingly promising tool of its own to study hadronic physics and high-energy physics~\cite{Silvi:2017srb,Banuls:2018jag}. Owing to its similarity with quantum computing, TN, despite being a classical computation approach, is used as a natural testbed for quantum simulation, especially for low dimensional and low entanglement problems. In particular, TN is an efficient simulation strategy to simulate quantum-many body systems by representing the quantum states locally in the Hilbert space and their dynamics can often be bounded by an area-law of entanglement entropy~\cite{Orus:2013kga,mps-reps-gs}. 
Importantly, TN is free from sign problems and reliable in accessing regimes inaccessible to Monte Carlo methods~\cite{Banuls:2018jag}. 
Recently, TN methods are successfully demonstrated in evaluations of hadronic vacuum polarization~\cite{Barata:2024bzk} and the PDF~\cite{Banuls:2024oxa,Banuls:2025wiq} in the 1+1 QED for the first time.
Here, we aim to compute both the PDF and DA in the same setting of the Nambu--Jona-Lasinio (NJL) model in 1+1 dimensions using efficient TN simulation algorithms. 
Having direct access to hundreds of qubits with TN, we present numerical simulation at large-qubit limit that is representative of the continuum limit. Specifically, we compare the behavior of the PDF and DA at various mass values and strong coupling strengths to make qualitative comparison with available results in perturbative QCD and the non-relativistic limits.

We organize this paper as follows. In Sect. II, we introduce the PDF and DA both in general and also in the context of the 1+1 dimension NJL model. In addition, we detail its mapping to qubits and introduce suitable TN algorithms. In Sect. III, we present and discuss numerical results obtained using TN simulation. In particular, we study the numerical convergence, mass/strong coupling dependence, and comparison with perturbative calculations. In Sect. IV, we summarize the paper and discuss possible future directions.

\section{Theoretical setup and computational methods}\label{sec:theory}

\subsection{Parton distribution functions and amplitudes}

Parton distribution functions (PDFs) are essential for describing inclusive processes at large momentum transfer in high-energy scattering, as dictated by the QCD factorization theorem~\cite{Collins:1989gx}. 
For example, the PDF for quarks within collinear factorization describes the probability for finding a quark carrying a longitudinal momentum fraction at some resolution scale.
These functions are fundamentally defined through the matrix elements of quark field operators~\cite{Collins:2011zzd}, providing a direct connection between partonic structure and scattering observables. The PDF can be expressed by,
\begin{align}
    \label{eq:pdf}
    f_{q/h}(x) =& \int_{-\infty}^{\infty} \frac{\mathrm{d}t}{4\pi} e^{-itx \vec{n}\cdot \vec{P}} 
    \bra{h(\vec{P})} \bar{\psi}(t\vec{n})W(t\vec{n}\leftarrow \vec{0}) \vec{n}\cdot\vec{\gamma} \psi(\vec{0}) \ket{h(\vec{P})} \,,
\end{align}
where \(x\) denotes the longitudinal momentum fraction carried by the quark of the hadron \(h(\vec{P})\) with momenta $\vec{P}$, \(\psi\) represents the quark field operator, and \(W(t\vec{n}\leftarrow \vec{0})\) is the Wilson line ensuring gauge invariance of the operator along the light-cone coordinate $\vec{n}$. In principle, one should expect the PDF satisfies,
\begin{align}\label{eq:mom_sumrule}
    \sum_{q_i} \int_0^1 \mathrm{d}x\, xf_{q_i/h}(x) = 1\,,
\end{align}
which is the momentum sum rule for all the partons.

On the other hand, the light-cone distribution amplitude (DA) governs exclusive processes at large momentum transfer~\cite{Lepage:1980fj,Efremov:1979qk}. It is formulated from the light-like vacuum-to-meson matrix elements. For a hadron \(h\), its DA is generally defined as the light-like separated, gauge-invariant vacuum-to-meson matrix element,
\begin{align}
    \label{eq:lcda}
    \phi_{h}(x) =& \frac{1}{f_h} \int_{-\infty}^{\infty} \mathrm{d}t\, \, e^{-it(x-1) \vec{n}\cdot \vec{P}} 
    \bra{\Omega} \bar{\psi}(t\vec{n})W(t\vec{n}\leftarrow \vec{0}) \vec{n}\cdot\vec{\gamma} \psi(\vec{0}) \ket{h(\vec{P})} \,,
\end{align}
where $f_h$ is the decay constant of the hadron, and $\ket{\Omega}$ represents the vacuum state. The DA encapsulates the distribution of partons inside the hadron along the light cone and plays a crucial role in describing exclusive processes within the framework of perturbative QCD. It is sometimes useful to compute the moments of the DAs, $\braket{\xi^n} = \int_0^1 dx \xi^n\phi(x)$ for $\xi=2x-1$, to compare with other approaches quantitatively. Notably, $\braket{\xi^0} = 1$ gives the normality of the DA.

Since both $f_{q/h}(x)$ and $\phi_{h}(x)$ are boost invariant, the computation of the PDF and DA can be conveniently carried out in the hadron rest frame. In this frame, where $\vec{P}=0$, the above expressions can be simplified. Specifically, one replaces the $|h(\vec{P})\rangle$ with $|h(0)\rangle$, and the momenta $\vec{P}$ with the mass $M_h$ of the specified hadron $h$ in the PDF and DA. Still, computational difficulty remains. In terms of the Hamiltonian matrix, both the hadron state preparation and the real-time evolution of the Wilson line and field operators inherently require exact diagonalization. The complexity for diagonalizing an $n$-dimensional Hamiltonian (not necessarily sparse) essentially scales as $\mathcal{O}(n^3)$ and therefore direct computation becomes a formidable if not an impossible task for a full size Hamiltonian that is relevant for comparison with experiments\footnote{Though computationally expensive, classical Hamiltonian calculations for hadrons that are in good agreement with experimental data do exist using light-front Hamiltonian in the leading Fock sectors with sparse diagonalization~\cite{Li:2017mlw,Lan:2019vui,Tang:2019gvn,Shuo_Bc,Qian:2020utg}.}. The goal of this work is to utilize tensor network methods as an alternative path for calculating the PDF and DA directly from first principles.

\subsection{Nambu--Jona-Lasinio model in 1+1 dimensions}

Simulating the (3+1)-d QCD Hamiltonian for hadronic structures is presently very challenging; therefore in this work we focus on the Nambu--Jona-Lasinio (NJL) model in 1+1 dimensions to demonstrate the computation of the PDF and DA in a common footing. The same procedure can also be applied to hadronic tensors. The NJL model is an effective model for low-energy two-flavored QCD~\cite{Nambu:1961tp,Nambu:1961fr} and its simplified and renormalizable version, the Gross-Neveu (GN) model~\cite{Gross1974}, is an ideal candidate for illustrative calculations including chiral symmetry breaking~\cite{Shi:2015ufa}, chiral condensate~\cite{Huang:2001yw}, phase transitions~\cite{Thies:2019ejd,Czajka:2021yll}, and so forth. We start with the Lagrangian density of the GN model with zero chemical potential, given by, 
\begin{align}
    \label{eq:Lagrangian}
\mathcal{L}&=\bar{\psi}(i\slashed{\partial}-m)\psi+g(\bar{\psi}\psi)^2\,,
\end{align}
where $m$ is the fermion mass and $g$ is the dimensionless coupling strength. Its Hamiltonian density is obtained via the Legendre transformation,
\begin{align}
\label{eq:Hamiltonian}
\mathcal{H}&=\bar{\psi}(i\gamma_1{\partial_1}+m)\psi-g(\bar{\psi}\psi)^2\,.
\end{align}
Here, the gamma matrices are chosen as $\gamma_0=\sigma^z$ and $\gamma_1=-i\sigma^y$, same as our previous work~\cite{Czajka:2021yll}. The fermion field $\psi(x)$ is discretized in position space and represented using staggered fermion fields~\cite{Borsanyi:2010cj} at each site $x_n = na$ on the lattice,
\begin{align}
\psi(x=x_n)=\left(\begin{array}{cc}
     &\ \rho(x=x_n)\  \\
     &\  \eta(x=x_n)\
\end{array}\right)=\frac{1}{\sqrt{a}}\left(\begin{array}{cc}
     &\ \chi_{2n}\  \\
     &\ \chi_{2n+1}\
\end{array}\right)\,,
\end{align}
where we use even $N$ for the total number of qubits and thus $\bar{N}=N/2$ for the total number of lattice sites. The lattice spacing is $\delta x=x_{n}-x_{n-1}=a$ where $n$ represents index of the lattice site. By using the forward-backward derivative convention~\cite{Czajka:2021yll, Chakrabarti:2003wi} for the partial derivative of the fermion field, we extract the full Hamiltonian in staggered fermion representation,
\begin{align}
\begin{split}
H=
&-\frac{i}{2a}\left[\sum_{n=0}^{N-2}\left(\chi_n^\dagger\chi_{n+1}-\chi_{n+1}^\dagger\chi_n\right)\right]+m\sum_{n=0}^{N-1}(-1)^n\chi_n^\dagger\chi_n\\
&-\frac{g}{a}\sum_{n=0}^{N/2-1}\left(\chi_{2n}^\dagger\chi_{2n}-\chi_{2n+1}^\dagger\chi_{2n+1}\right)^2\,.\label{Hamiltonia_all}
\end{split}
\end{align}
Here, we omit the periodic boundary terms $\chi_{N-1}^\dagger\chi_{0}-\chi_{0}^\dagger\chi_{N-1}$ in the kinetic term. In large systems, the long-range term has limited contribution and is complicated in the quantum simulation. To map the Hamiltonian onto quantum circuits, we apply the Jordan-Wigner (JW) transformation to the staggered fermion fields~\cite{Jordan:1928wi}, 
\begin{align}
\chi_n=\sigma_{n}^-\prod_{i=0}^{n-1}\left(-i\sigma_{i}^z\right), \quad \chi^\dagger_n=\sigma_{n}^+\prod_{i=0}^{n-1}\left(i\sigma_{i}^z\right) \,,\label{eq:jw-trans}
\end{align}
and obtain the spin Hamiltonian,
\begin{align}\label{eq:spin_Ham}
\begin{split}
    H&=\frac{1}{2a}\left[ \sum_{n=0}^{N-2}\left(-\sigma^+_n\sigma^z_n\sigma^-_{n+1} + \sigma^+_{n+1}\sigma^z_n\sigma^-_{n}\right)\right] + m\sum_{n=0}^{N-1}(-1)^n\sigma^+_n\sigma^-_n \\
    &- \frac{g}{a}\sum_{n=0}^{N/2-1}\left( \sigma^+_{2n}\sigma^-_{2n} - \sigma^+_{2n+1}\sigma^-_{2n+1}\right)^2\,,
\end{split}
\end{align}
where $\sigma_i^x, \sigma_i^y, \sigma_i^z$ are the Pauli matrices and $\sigma^{\pm}_i = (\sigma_i^x \pm i \sigma_i^y)/2$ on qubit $i$. 

Now we compute the PDF in eq.~\eqref{eq:pdf} and DA in eq.~\eqref{eq:lcda} using the obtained (1+1)-dimensional NJL Hamiltonian. Specifically, we use $\gamma^{\pm}= \frac{1}{\sqrt2}(\gamma^0 \pm \gamma^1)=\frac{1}{\sqrt2}(\sigma^z \pm i\sigma^y)$ and $\vec{n}=(1,1)$ for the light-cone coordinate.
It follows that $\vec{n}\cdot{\vec{\gamma}}=\sqrt{2}\gamma^-$ and $\vec{n}t = (1,1)z$ with a reparameterized variable $z$ for space and time. Since the gauge fields are absent in the NJL model, the explicit Wilson lines in the PDF and DA vanish and we are left with time evolution operators of the Hamiltonians. While this feature of the NJL model dramatically simplifies the calculation, we are well aware that one should include dynamical gluon contributions in a more realistic scenario with available methods for simulating the Wilson line~\cite{Zohar:2012ts,Brennen:2015pgn,Lamm:2018siq, Zohar:2019tsa,Echevarria:2020wct}, which we leave for a subsequent investigation. Recently, treatment of evolution of the Wilson lines was proposed and detailed for the Schwinger model~\cite{Banuls:2025wiq}.
Applying the same JW encoding to field operators $\psi^\dagger$ and $\psi$, the final expression of the PDF and DA are related to the sum of four matrix element expectations,
\begin{align}
    \label{eq:qc_pdf}
	f_{q/h}(x) =& \frac{1}{4\pi} \sum_{z=\bar{N}_\mathrm{min}}^{\bar{N}_\mathrm{max}} e^{-ix M_hz} \bra{h}\big(M_{00}(z)+M_{11}(z)-M_{01}(z)-M_{10}(z)\big)\ket{h}\,,\\
    \label{eq:qc_lcda}
	\phi_{h}(x) =&\frac{1}{f_h}\sum_{z=\bar{N}_\mathrm{min}}^{\bar{N}_\mathrm{max}}  e^{-i(x-1) M_hz} \bra{\Omega} \big(M_{00}(z)+M_{11}(z)-M_{01}(z)-M_{10}(z)\big)\ket{h}\,,
\end{align}
where each matrix element operator is defined by,
\begin{align}\label{eq:matrix_elements}
    M_{ij}(z) &= e^{iHz}\chi^{\dagger}_{2z+i+\bar{N}}e^{-iHz}\chi_{j+\bar{N}}\,,
\end{align}
with evolution time $z=na/2$ relating to sites $n$ and $n+1$, and $\bar{N}_\mathrm{min}$ and $\bar{N}_\mathrm{max}$ mark the boundary of the lattice. Here, we always center the system around $x=0$ positions for the simulation, so $\bar{N}_\mathrm{min}=-\lfloor \bar{N}/2 \rfloor$ and $\bar{N}_\mathrm{max}=\lfloor \bar{N}/2\rfloor$. We then have that qubits $\bar{N}, \bar{N}+1$ simulate $x = 0$. Then, for a sufficiently large $\bar{N}$, our setup is a representation of the physical system. In practice, when calculating the PDF, we also subtract the vacuum expectation from the hadron expectation due to finite system size used in our simulation~\cite{Collins:2011zzd}. 

\subsection{Tensor network methods}
Quantum computing technology emerges as one promising strategy to overcome classically difficult problems. Real-time evolution takes advantage of the trotter formula~\cite{nielsen_chuang_2010}, in principle providing an efficient way to compute the PDF and DA. Various quantum algorithms for obtaining the ground state and excited states also provide solid approaches to state preparation~\cite{wiersema2020exploring,Choi:2020pdg,Chakraborty:2020uhf,Lin:2020zni,Dong:2022mmq,Kane:2023jdo}. The final observables of interests are essentially expectation values that can be evaluated on quantum circuits via direct measurement and the Hadamard test~\cite{nielsen_chuang_2010}. In this work, we focus on the calculation of the PDF and DA in the context of tensor network (TN) methods that share many similarities with quantum computation.

Tensor networks provide a convenient alternative for many-body simulations, especially since today's quantum computers are still limited both in the number and fidelity of available qubits. TN methods recast (either exactly or up to accurate approximation) many-body quantum systems as a matrix product state (MPS)~\cite{mps-reps-gs} and use various algorithms and variational methods for studying the dynamics of MPS systems. MPS is particularly well-suited for representing the ground states of local, gapped Hamiltonians in 1D systems with a unique ground state and finite interaction strengths, where entropy area laws impose an upper bound on entanglement~\cite{Hastings:2007iok,vidal-erg}.
For this reason, TN methods provide efficient many body quantum simulations with classical hardware for calculating hadronic structures in the 1+1 dimensional model considered in this work\footnote{Note, for the same reason, TN do not replace standard QC methods but complement them.}.

\subsubsection{Density matrix renormalization group}
Preparation of the ground state or excited states is often the starting point for any quantum simulation~\cite{Jordan:2011ci}. While many quantum simulation methods have been suggested, including adiabatic quantum computation~\cite{Chakraborty:2020uhf}, Hamiltonian variational algorithm~\cite{wiersema2020exploring}, quantum subspace expansion~\cite{mcclean2017hybrid}, Rodeo algorithm~\cite{Choi:2020pdg}, among others; it is still typically very complicated to prepare the states on quantum circuits. In TN, by comparison, this complication is largely reduced utilizing efficient representation of the Hilbert space using MPS. 
An MPS is an ansatz for the many-body wave function composed of local tensors,
\begin{equation}\label{eq:mps}
    \ket{\psi} = \sum_{s_1 s_2,...,s_N}^{}  A_{1}^{s_1}A_{2}^{s_2}\cdots A^{s_N}_{N} \ket{ s_1 s_2 \cdots s_N}\;.
\end{equation}
Each $A_i^{s_i}$ is a matrix associated with site $i$ and state $s_i$ with size $\chi \times \chi$ where $\chi$ is the bond dimension between site $i$ and $i+1$. For open boundary conditions, $A_1^{s_1}$ is a $1 \times \chi$ row vector and $A_N^{s_N}$ is a $\chi \times 1$ column vector. The bond dimension $\chi$ determines the entanglement capacity; a larger $\chi$ captures more correlations but increases computational cost.

In this work, we use a variational ground-state search eigensolver to calculate the lowest-energy eigenvector and eigenvalue of the system with respect to the two-site block by variationally minimizing the energy expectation value $\braket{\psi |H | \psi}/\braket{\psi | \psi}$ of a given Hamiltonian $H$ to obtain the ground state or first excited state $\ket{\psi}$ in MPS form~\cite{White:1992zz,White:1993zza,Schollwock2005,schollwock2011density,frank-dmrg} with the ITensor library~\cite{ITensor:2022}. This algorithm takes the MPS chain and holds all tensors fixed except for two neighboring sites, $A_{i}$ and $A_{i+1}$. An eigensolver calculates the lowest-energy eigenvector and eigenvalue of the system with respect to the two-site block. Afterwards, the two-site tensor is split back into two separate sites via singular value decomposition, the new bond dimension between the two sites can be defined as higher than the original bond dimension. This splitting allows the bond dimension to dynamically grow with evolution and the process is repeated for different sites until convergence is reached. For a detailed description and comparison, see Refs.~\cite{schollwock2011density,ostlund1995thermodynamic}.

Traditionally, the variational algorithm is used to study ground states of one-/two-dimensional many body systems~\cite{frank-dmrg}. The computational time scaling is $\mathcal{O}(N \chi^3)$, where $\chi$ is bond dimension and $N$ is system size, i.e., total number of qubits~\cite{dmrg-on-tpul}. 
Specifically, the bond dimension is the dimension of the connecting index between two local tensors of an MPS.
For calculating the first excited state, we construct a modified Hamiltonian that consists of the original Hamiltonian and a projection of the previously calculated ground state wave function. The projection has a coefficient ``penalty" term that penalizes overlaps between the excited and ground states, which ensures orthogonality and minimization outside of the ground state~\cite{ITensor:2022}. In this way, we use the variational algorithm to obtain both the ground state and the first excited state as MPS for the NJL model. 

\subsubsection{Time dependent variational principle}
To calculate the matrix elements for the PDF and DA, as in eq.~\eqref{eq:matrix_elements}, we evolve the prepared MPS state in real time and collapse with a final MPS. The two primary algorithms for the time evolution of the MPS are the time-evolving block decimation (TEBD)~\cite{Vidal:2003lvx,verstraete2004matrix} and time dependent variational principle (TDVP)~\cite{Haegeman:2011zz, haegeman2016unifying}. Both of them have the same computational time scaling of $\mathcal{O}(N \chi^3)$ as the variational search algorithm, where $\chi$ is the bond dimension and $N$ is the system size. For time step $\tau$, the TDVP and TEBD algorithms have $O(\tau^2)$ error scaling for the lower order~\cite{dmrg-on-tpul,tdvp-scaling,vidal-tebd-many-body}. 

In principle, the TEBD method breaks up the Hamiltonian into small chunks and applies real or imaginary time evolution via Trotter gates that account for near neighbor interactions~\cite{vidal-tebd-qc,vidal-tebd-many-body}.
The entanglement, in the worst case, can grow linear in time, which means the bond dimension of the MPS has to grow exponentially over time for a faithful representation of the state~\cite{Kim:2013etb,Schuch_2008, Calabrese:2005in, Schachenmayer:2013}. In general, the entanglement growth is highly problem dependent. For gapped systems that evolve adiabatically close to the ground state at any time, typically it does not lead to a significant growth of entanglement, while out-of-equilibrium dynamics can trigger the worst case scenario~\cite{Orus:2013kga}.
If the bond dimension is not allowed to grow to account for this, then truncation errors occur~\cite{vidal-tebd-qc,vidal-tebd-many-body}.

On the other hand, the TDVP algorithm projects the MPS onto a tangent space manifold, on which it solves the time dependent Schrodinger equation. TDVP solves the Schrodinger equation projected onto a manifold, often chosen via the global Krylov method, which globally maintains all interaction terms \cite{Haegeman:2011zz}. Potential errors are not as controllable, but are reduced with the use of Krylov time evolution \cite{itensor-tdvp}. 
Though the TDVP and TEBD algorithms have the same scaling up to a prefactor, the TDVP algorithm has both a smaller prefactor for computational time scaling and less bond dimension growth~\cite{Haegeman:2011zz,itensor-tdvp}. In this work, we primarily used TDVP, as it showed better performance than TEBD for NJL model, espeicially for high qubit simulations.

\section{Results}\label{sec:result}
In this work, we prepare the hadron state and simulate the real-time correlators of the quark and gluon fields using tensor network framework~\texttt{ITensor}~\cite{ITensor:2022}. The numerical results are compared with available exact diagonalization and quantum simulation results. Throughout this paper, we always use the lattice spacing $a=1$ in our simulations. Importantly, our simulation follows the implementation using variational ground-state search algorithm for state preparation and the TDVP for time evolution; whereas the recent work on Schwinger model uses TEBD for the time evolution~\cite{Banuls:2025wiq}.

\subsection{Hadron state preparation}

Here, within the 1+1 dimension NJL model, our hadronic states are prepared via two-site varitional ground-state search algorithm. To prepare the hadron state on a tensor network within the neutral charge $Q=0$ sector, mimicking a charge-neutral pion ($\pi^0$), we utilize a charge-preserving (CP) lattice to conserve the quantum number. With our setup, the total charge operator is defined as $Q=\sum_{i=1}^N (\sigma^z_n+(-1)^n)/2$ on a lattice of $N$ qubits~\cite{Ikeda:2024rzv}. When applying variational algorithms to obtain both the vacuum state and hadron state, we start with a guess ansatz that we know analytically has $Q = 0$; then the ground-state search algorithm ensures that we remain in the $Q = 0$ sector. It is important to note that when working on a CP lattice, the Hamiltonian operator should be written in a form where its terms consist of only $\sigma^z$ or both $\sigma^-, \sigma^+$ to ensure charge conservation, as in eq.~\eqref{eq:spin_Ham}. We first obtain the vacuum state (i.e. the ground state) on a CP lattice with sufficient number of sweeps and numerical accuracy. Specifically, we found 100 sweeps and 200 {maximum bond dimensions} sufficient for simulating a system of 100 qubits at a cutoff of $10^{-12}$. By sufficient we mean that the bond dimension was not fully exhausted during the variational search. The high sweeps are found to be crucial to ensure numerical convergence for simulation at larger qubits. 

To obtain the hadron state (i.e., the first excited state), we project out the vacuum MPS and perform another set of sweeps with sufficiently large weights. For the hadron state, we found that even higher sweeps and maximum bond dimension are necessary to reach convergent mass results. Specifically, we used up to 500 sweeps and 800 maximum bond dimension. Throughout our simulation, we ensured that the obtained MPS consistently maintained a total charge of zero and that the bond dimension remained well below the allocated maximum limit. In Table.~\ref{tab:dmrg}, we summarize our hadron mass $M_h$, defined as the mass difference between the excited state and the ground state, findings for various system sizes $N$:
\begin{align}
    M_h = \langle h| H|h\rangle - \langle \Omega|H|\Omega\rangle\,.
\end{align}
Specifically, each value of $N$ is picked such that the two boundaries are matched up, i.e., $\bar{N}_\mathrm{max}=-\bar{N}_\mathrm{min}$. 
As the system size increases, $M_h$ becomes stable and we are able to identify the mass gap with high accuracy. For this reason, we always use $N=102$ qubits in preparing the ground and excited states for the final results. Note the hadron mass values in Table.~\ref{tab:dmrg} are converged results obtained with sufficient maximum bond dimension with a cutoff of $10^{-12}$. See Table.~\ref{tab:dmrg_details} in Appendix 1 for a detailed summary of the hadron mass values at various numerical cutoffs and maximum bond dimensions that shows our results converge with increasing maximum bond dimensions and cutoffs, respectively.

\begin{table}[htp!]
\setlength\tabcolsep{8pt}
 \centering
 \caption{Hadron masses obtained from variational ground-state search algorithm on CP lattice for different qubits. Here, $m=0.7a^{-1}$ and $g=0.25$. The results are accurate and converged with sufficient sweeps (100-500) and maximum bond dimensions (200-800) at a cutoff $10^{-12}$ for each qubit size.} 
 \label{tab:dmrg} 
 \begin{tabular}{ |c|  c|  c|  c| c| c|} %m{3cm} m{3cm}
\hline
 Qubits (N)
 & $18$
 & $30$
 & $50$
 & $102$
 \\
 \hline
 Hadron mass $M_h a$ & 1.64600864 & 1.64359027 & 1.64246829 & 1.64187901 \\ 
 \hline
 \end{tabular}
\end{table}

\subsection{Real-time evolution}
% \WQ{@Peter, add PDF dependence on cutoff for maximum z values}
To compute the expectation of each matrix element as in eq.~\eqref{eq:matrix_elements} in the PDF and DA, we apply the local operators, then real-time evolve the obtained MPS, and lastly collapse on the other MPS state to compute their overlap. Essentially, we repeatedly evaluate the following expectation,
\begin{align}\label{eq:matrix_element_expectation}
    E_{ab}(z) &= \braket{\psi_L | e^{iHz}\chi^{\dagger}_{a}e^{-iHz}\chi_{b} | \psi_R}\equiv e^{iE_Lz}\braket{\psi_L | \chi^{\dagger}_{a}e^{-iHz}\chi_{b} | \psi_R}\,,
\end{align}
where $\ket{\psi_L}, \ket{\psi_R}$ can be either $\ket{h}$ or $\ket{\Omega}$. Since $\ket{\psi_L}$ is an eigenstate of $H$ (with eigenvalue $E_L$), the left-side time evolution operator reduces to an overall phase factor that greatly simplifies the computation\footnote{We verified numerically the two expressions for $E_{ab}(z)$ in eq.~\eqref{eq:matrix_element_expectation} yield identical results.}. Applying the local operators on MPS is straightforward and directly built into \texttt{ITensor} that involves three designated procedures: contraction, unpriming, and insertion of the MPS tensor.
To time evolve the MPS, we use the TDVP algorithm~\cite{Haegeman:2011zz,haegeman2016unifying}, included with the \texttt{ITensorTDVP} library. 
Specifically, we always use one time sweep and a maximum bond dimension of 810 (slightly larger than our max hadron state bond dimension). We tested the effect of different number of TDVP sweeps on the simulation results and found no significant improvement with more than one sweep. For large system sizes, we note that both the bond dimension of the MPS state and cost of the simulation increase at longer evolution time $z$. To avoid hitting the allocated maximum bond dimension and reduce computational complexity, we simulate in the region between $z=-10$ and $z=10$ where the matrix elements have substantial non-zero contributions. We verified that simulating beyond this region yielded no significant improvement to our final results.

\subsection{Parton distribution function and amplitudes}

With our prepared hadron state and real-time evolution, we extract expectations of hadron matrix elements and then Fourier transform to obtain the PDF in eq.~\eqref{eq:qc_pdf} and DA in eq.~\eqref{eq:qc_lcda}. Specifically, after obtaining the matrix element sums for various $z$ values, we interpolate them using cubic polynomial interpolator from \texttt{scipy} to generate a smooth curve prior to their Fourier transform (FT). 
For the FT, we used both the discrete and continuous FTs where we find no dramatic difference between them. For the numerical results presented below, we always show the continuous FT calculated using the composite Simpson's rule from \texttt{scipy}.
In addition, we use the momentum sum rule of the PDF and zero moment of the DA to normalize our obtained distributions after the FT respectively.
In the case of the PDF, we also subtract the vacuum expectation of the matrix elements from the hadronic expectation before performing the FT. 

\begin{figure}[thp!]
    \centering
    \subfigure[\;PDF calculated using different methods\label{fig:comparison}]{\includegraphics[width=0.47\textwidth]{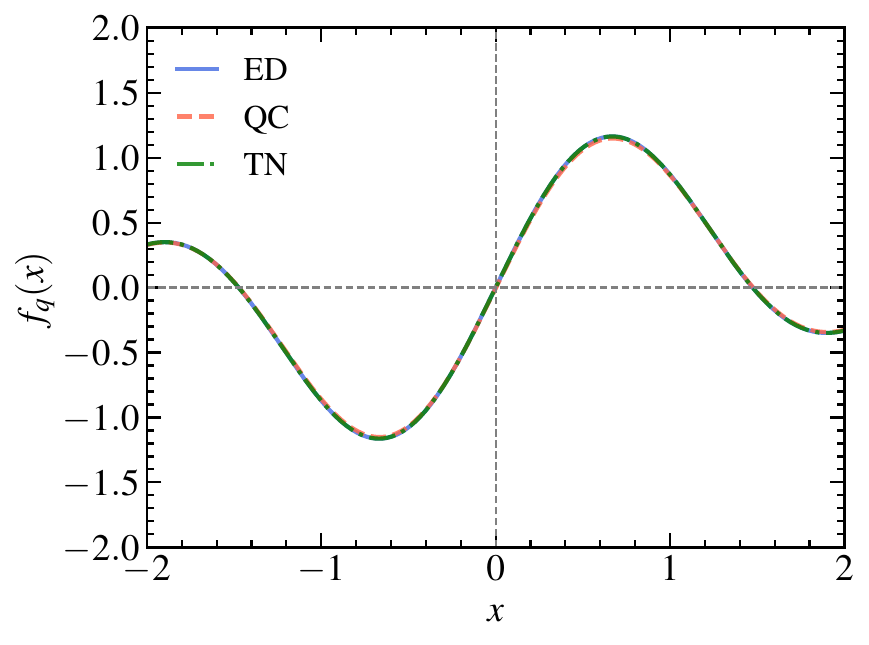}
    }\quad
    \subfigure[\;Errors in PDF compared to ED\label{fig:comparison}]{\includegraphics[width=0.47\textwidth]{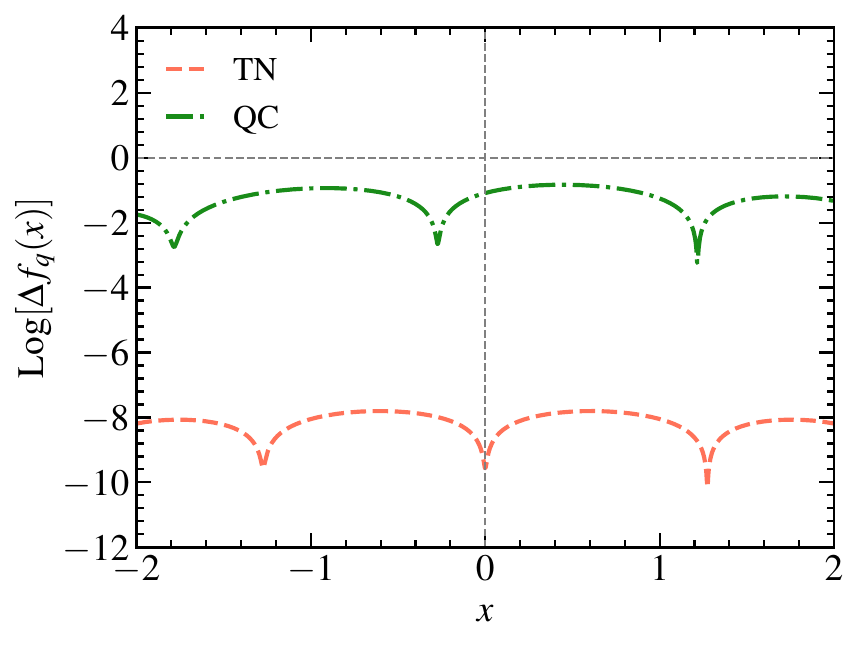}
    }
    \caption{Simulation results for the PDF comparing tensor network, exact diagonalization, and quantum computing methods. We use $N = 10$, $m=0.7a^{-1}$, and $g=0.25$. Right plot shows the error of the PDF calculated using tensor network and quantum circuit simulation.
    }
\label{fig:method_comparison}
\end{figure}

For small system sizes, we find consistent agreement among the TN approach, exact diagonalization (ED), and quantum circuit (QC) simulation. For the quantum circuit simulation, we used first-order trotterization~\cite{Trotter1959} to implement time evolution with scaling $\mathcal{O}(\tau^2)$ for the time step $\tau$ that matches the order of TDVP~\cite{childs2019nearly, childs2021theory}. Specifically, we use the statevector simulator from {\texttt Qiskit}~\cite{Qiskit} library to perform exact simulation with no sampling noise from shots.
State preparations for the QC simulation is dramatically simplified and taken as exact eigenstates directly solved from diagonalization. In Fig.~\ref{fig:method_comparison}, we present the numerical results for the PDF using a lattice of $N=10$ qubits. From Fig.~\ref{fig:method_comparison}(a), we see a strong alignment among all three results, implying that our setup is valid and generally applicable to quantum computing. In Fig.~\ref{fig:method_comparison}(b), we make a log-linear plot of the absolute error $\Delta f_q(x)$ in TN and QC methods compared with exact diagonalization for each $x$. Noticeably, the TN consistently and significantly reduces the error  by six orders of magnitude with an average error of around $10^{-8}$, which aligns with the numerical accuracy of the variational method for state preparation. Similar observations are found in computing the DA at $N=10$ qubits. These observations prove the reliability of the TN strategies used in our numerical calculations.

\begin{figure}[thp!]
    \centering
    \subfigure[\;PDF\label{fig:PDF_vs_N}]{\includegraphics[width=0.47\textwidth]{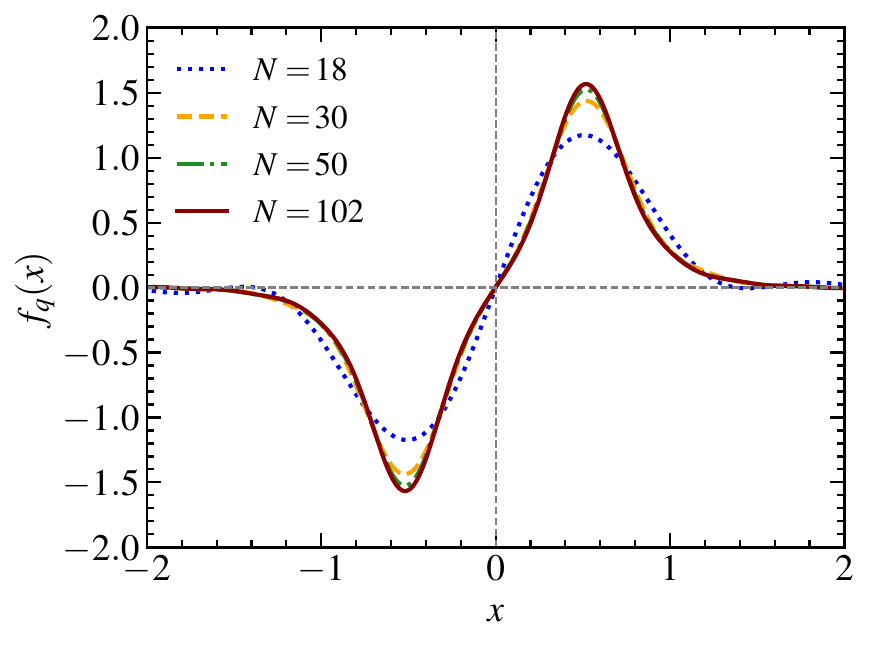}
    }\quad
    \subfigure[\;DA\label{fig:LCDA_vs_N}]{\includegraphics[width=0.46\textwidth]{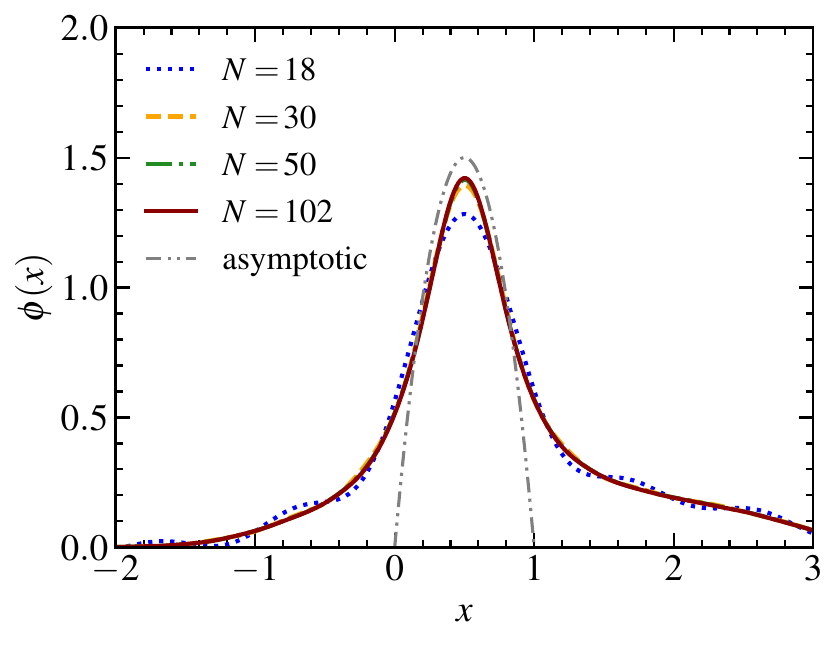}
    }
    \caption{Simulation results for the PDF and DA over increasing number of qubits $N$. Here, we fix $m=0.7a^{-1}$ and $g=0.25$. The asymptotic of perturbative QCD for the DA is provided.
    }
\label{fig:vs_N}
\end{figure}

To demonstrate numerical results in the continuum limit, we use sufficiently high number of qubits in our simulations to compute the PDF and DA using tensor networks. Such calculation is beyond the current limits of exact simulation and quantum simulation with classical hardware. With tensor networks, we are able to study physical results at continuum limits utilizing an efficient representation of the many-body system with the MPS. In Fig.~\ref{fig:vs_N}, we compute both the PDF and DA in the same setting with increasing system sizes (i.e., qubits) using selected mass $m=0.7a^{-1}$ and strong coupling $g=0.25$. 
For the PDFs, as expected in a one-flavor hadron (such as $\pi^0$, composed of many pairs of $q$ and $\bar{q}$), we see an asymmetric plot with a peak at $x = 0.5$ and $x=-0.5$.
Similarly for the DAs, we see a typical distribution centered in the $x\in[0,1]$ region.
As system size $N$ increases, both the PDF and DA results gradually converge to the large-qubit curve at $N=102$, suggesting the calculations approach the continuum of $N\rightarrow \infty$ limit\footnote{In principle, one would take both $a\rightarrow 0$ and $N\rightarrow \infty$. For simplicity, we fix $a=1$.}. Therefore, for all the results below, we always use $N=102$ qubits for the simulation and we find it sufficient to represent the continuum behavior of the PDF and DA at this large qubit limit for our NJL Hamiltonian. Overall, the shape and tendency of our results are qualitatively consistent with experimental observations such as E791 data~\cite{E791:2000xcx} for the PDF and Fermilab E615 data~\cite{Conway:1989fs} for the DA of the pion and they also resemble numerical calculations in 1+1 QCD~\cite{Jia:2018qee}.

\begin{figure}[thp!]
    \centering
    \centering
    \subfigure[\;PDF\label{fig:pdf_fix_m}]{\includegraphics[width=0.47\textwidth]{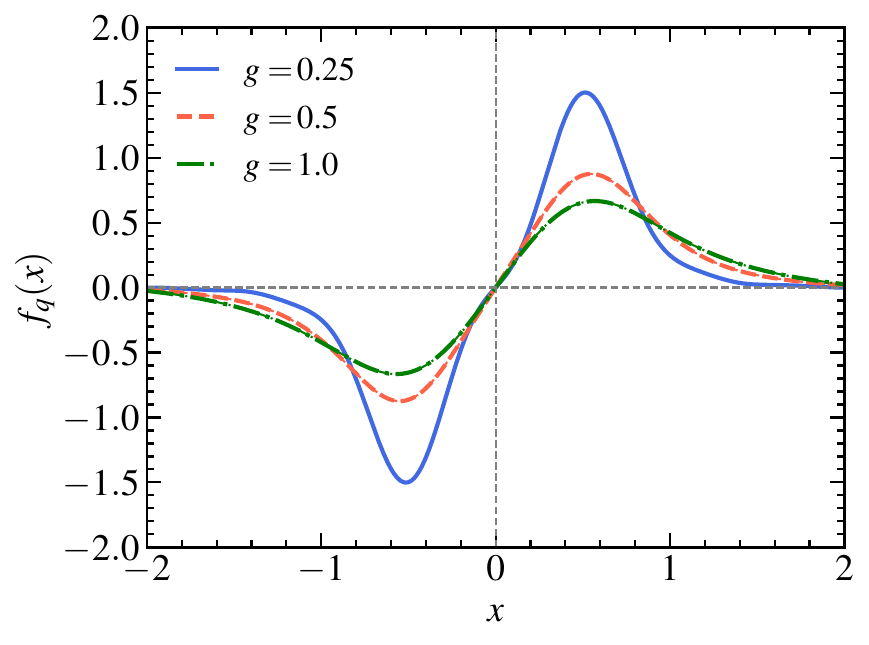}
    }\quad
    \subfigure[\;DA\label{fig:lcda_fix_m}]{\includegraphics[width=0.46\textwidth]{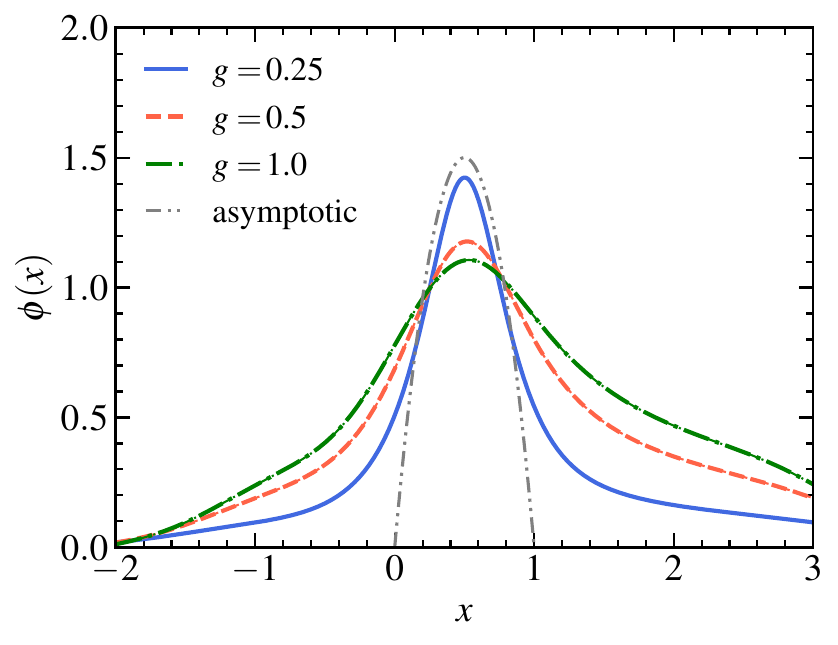}
    }
    \caption{Simulation results for the PDF and DA at different values of strong coupling $g$. The mass is fixed at $m=0.5a^{-1}$ and total qubits are $N=102$.
    }
\label{fig:fixed_mass}
\end{figure}

One important strength of our approach is the freedom to investigate the PDF and DA at various coupling strengths, especially in the non-perturbative regions that are generally complicated. In Fig.~\ref{fig:fixed_mass}, we show the PDF and DA results at various coupling strengths $g=0.25,0.5,1.0$ with the same $m=0.5a^{-1}$ and $N=102$ qubits. Interestingly, the PDFs show distinctive behavior at different coupling strengths. At small coupling $g=0.25$, we get a sharp peak at $x=0.5$ and $x=-0.5$ that is consistent with the perturbative QCD expectation. With increased coupling strength, the PDF becomes less sharply peaked. Similarly for the DA, at the low coupling such as $g=0.25$, the distribution is sharply centered at $x=0.5$, comparable to the perturbative limit of $\phi_{\mathrm{asym}}(x) = 6x(1-x)$~\cite{Lepage:1980fj}. When we turn on the coupling strength towards $1$, we see significant non-perturbative effects, giving broadened DA curves. Qualitatively, our results are also in agreement with available quantum and tensor network simulation results from the light-front Hamiltonian methods~\cite{Kreshchuk:2020dla,Kreshchuk:2020aiq,Qian:2021jxp} and lattice calculations~\cite{LiTianyin:2021kcs,LiTianyin:2024nod,Banuls:2024oxa,Banuls:2025wiq}.

\begin{figure}[thp!]
    \centering
    \centering
    \subfigure[\;PDF\label{fig:pdf_fix_m}]{\includegraphics[width=0.47\textwidth]{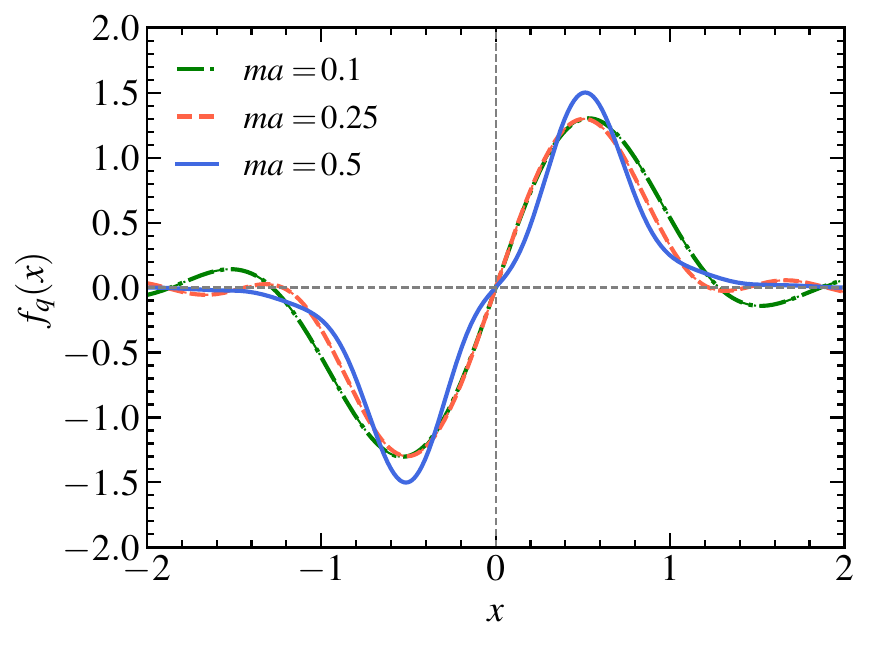}
    }\quad
    \subfigure[\;DA\label{fig:lcda_fix_m}]{\includegraphics[width=0.46\textwidth]{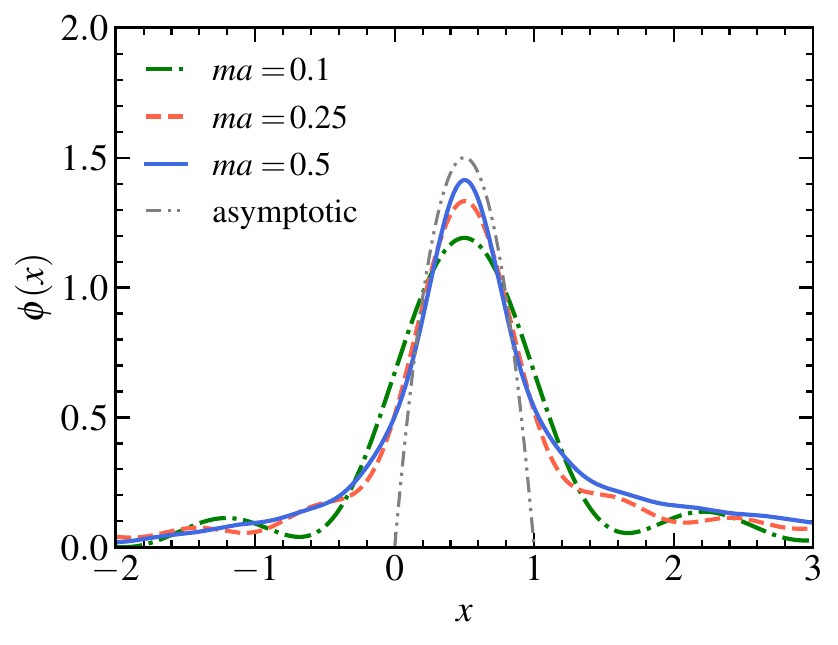}
    }
    \caption{Simulation results for the PDF and DA at different values of mass $m$. The strong coupling is fixed at $g=0.25$ and total qubits are $N=102$.
    }
\label{fig:fixed_coupling}
\end{figure}

Likewise, we compute the PDF and DA for various mass values which may serve to illustrate the separated behavior between heavy systems and light systems. In Fig.~\ref{fig:fixed_coupling}, we compute the PDF and DA at masses $m=0.1a^{-1}, 0.25a^{-1}, 0.5 a^{-1}$ with fixed $g=0.25$ and $N=102$ qubits. In both the PDF and DA, we find a more peaked distribution at $x=0.5$ for the heavier masses and broadened distributions for the lighter masses. We also observe less oscillations on the tail ends of the distributions as mass decreases. Importantly, our findings are comparable to the light-front Hamiltonian approach~\cite{Brodsky:1997de,Brodsky:2014yha,1stBLFQ} for mesonic systems of distinctive masses, from light mesons~\cite{Vega:2009zb,Swarnkar:2015osa,Lan:2019vui,Jia:2018ary,Qian:2020utg} to heavy quarkonia~\cite{Swarnkar:2015osa,Li:2017mlw,Lan:2019img}, where leading Fock sectors are used. It would be interesting to further address the large-qubit calculations for these different physical systems by including flavor degrees of freedom into our Hamiltonian.

\section{Conclusion and outlook}\label{sec:conclusion}
In this work, we used a quantum strategy based on tensor network simulation to extract the PDF and DA for hadrons in the Nambu--Jona-Lasinio model in 1+1 dimensions in the same setting. Our approach utilizes the variational ground-state search algorithm to accurately construct the hadron and vacuum states on a charge-preserving lattice and the TDVP algorithm to efficiently time evolve the two-point correlator needed for these non-perturbative quantities. With our approach, we extract the PDF and DA in the same setting with converging results up to 102 qubits. We are able to study the behavior of the PDF and DA under the same setting for various masses and strong couplings, and observe consistent agreement with the expectations from the perturbative QCD as well as non-relativistic limit. Our exploratory study shows the advantage of tensor network in the current era for computing various physically important quantities in the continuum limit.

It would be very important to expand non-perturbative calculations of quantities such as the PDF, DA, and hadronic tensors using tensor network to U(1) and SU(2) gauge theories. Notably, similar calculation using TN methods has been successfully demonstrated in Ref.~\cite{Banuls:2025wiq} for the U(1) case in the Schwinger model. Inclusion of multiple flavor and spin degrees of freedom in the line of study would provide significant understanding to realistic mesons and baryons from first principles. While tensor networks have primarily been applied to 1+1 dimensions, it would be interesting to extend to higher lattice dimensions. The natural extension of MPS to higher dimensions would be projected entanglement pair states~\cite{Verstraete:2004cf,verstraete2004valence}, tree tensor network~\cite{Shi:2005ctp}, and so forth, which have been used to study 3+1 dimension Lattice Gauge Theory~\cite{Magnifico:2020bqt, Banuls:2018jag}.
We also plan to incorporate more advanced simulation techniques, such as tensor renormalization group~\cite{LevinNave:2007} and its natural extension, tensor network renormalization~\cite{tnr}, to scale to higher qubit simulations with reduced error and study entanglement structure. These methods can be used to calculate correlation functions and generalize to systems beyond one spatial dimension.

\section*{Acknowledgement}
We are grateful to Jo\~ao Barata, Kazuki Ikeda, Tianyin Li, Yang Li, David Lin, Manuel Schneider, and Hongxi Xing for their helpful and valuable discussions. ZK, NM, and PN are supported by National Science Foundation under grant No.~PHY-1945471. WQ is supported by the European Research Council under project ERC-2018-ADG-835105 YoctoLHC; by Maria de Maeztu excellence unit grant CEX2023-001318-M and project PID2020-119632GB-I00 funded by MICIU/AEI/10.13039/501100011033; by ERDF/EU; by the Marie Sklodowska-Curie Actions Fellowships under Grant No.~101109293; and by Xunta de Galicia (CIGUS Network of Research Centres).

\appendix
\section{Hadron mass details obtained from tensor network simulations}
\begin{table}[htp!]
\setlength\tabcolsep{8pt}
 \centering
 \caption{Hadron masses obtained from variational ground-state search algorithm on CP lattice for different qubits with various truncation error cutoffs $\Delta$ and maximum bond dimensions $D_\mathrm{max}$. Here, $m=0.7a^{-1}$ and $g=0.25$ are used. For varying cutoffs, we use fixed $D_\mathrm{max}=800$; for varying maximum bond dimensions, we use fixed $\Delta=10^{-12}$. All simulations are done with sufficient number of sweeps (100-500) depending on system sizes. 
 }
 \label{tab:dmrg_details} 
 \begin{tabular}{ |c|  c|  c|  c| c| c|} %m{3cm} m{3cm}
\hline
 Qubits (N)
 & $18$
 & $30$
 & $50$
 & $102$
 \\
 \hline\hline
 $\Delta=10^{-6}$  & 1.64599824 & 1.64356479 & 1.64242277 & 1.64187559 \\ \hline 
 $\Delta=10^{-8}$  & 1.64600860 & 1.64359023 & 1.64246818 & 1.64188719 \\ 
 \hline
 $\Delta=10^{-10}$ & 1.64600864 & 1.64359027 & 1.64246829 & 1.64188740 \\ 
 \hline
 $\Delta=10^{-12}$ & 1.64600864 & 1.64359027 & 1.64248883 & 1.64188740 \\ 
 \hline
\hline
 Qubits (N)
 & $18$
 & $30$
 & $50$
 & $102$
 \\
 \hline\hline
 $D_\mathrm{max}=40$  & 1.64600864 & 1.64359029 & 1.64246831 & 1.64188746 \\ \hline 
 $D_\mathrm{max}=80$  & 1.64600864 & 1.64359027 & 1.64248690 & 1.64188740 \\ 
 \hline
 $D_\mathrm{max}=100$ & 1.64600864 & 1.64359027 & 1.64248918 & 1.64188740 \\ 
 \hline
 $D_\mathrm{max}=200$ & 1.64600864 & 1.64359027 & 1.64248883 & 1.64188740 \\ \hline 
 $D_\mathrm{max}=400$ & 1.64600864 & 1.64359027 & 1.64248883 & 1.64188740 \\ \hline 
 $D_\mathrm{max}=800$ & 1.64600864 & 1.64359027 & 1.64248883 & 1.64188740 \\ 
 \hline
 \end{tabular}
\end{table}

\bibliographystyle{JHEP}
\bibliography{main.bib}

\providecommand{\href}[2]{#2}\begingroup\raggedright\begin{thebibliography}{100}

\bibitem{AbdulKhalek:2021gbh}
R.~Abdul~Khalek et~al., \emph{{Science Requirements and Detector Concepts for the Electron-Ion Collider}: {EIC Yellow Report}}, \href{https://doi.org/10.1016/j.nuclphysa.2022.122447}{\emph{Nucl. Phys. A} {\bfseries 1026} (2022) 122447} [\href{https://arxiv.org/abs/2103.05419}{{\ttfamily 2103.05419}}].

\bibitem{Accardi:2012qut}
A.~Accardi et~al., \emph{{Electron Ion Collider: The Next QCD Frontier}: {Understanding the glue that binds us all}}, \href{https://doi.org/10.1140/epja/i2016-16268-9}{\emph{Eur. Phys. J. A} {\bfseries 52} (2016) 268} [\href{https://arxiv.org/abs/1212.1701}{{\ttfamily 1212.1701}}].

\bibitem{Collins:1989gx}
J.C.~Collins, D.E.~Soper and G.F.~Sterman, \emph{{Factorization of Hard Processes in QCD}}, \href{https://doi.org/10.1142/9789814503266_0001}{\emph{Adv. Ser. Direct. High Energy Phys.} {\bfseries 5} (1989) 1} [\href{https://arxiv.org/abs/hep-ph/0409313}{{\ttfamily hep-ph/0409313}}].

\bibitem{Ji:2013dva}
X.~Ji, \emph{{Parton Physics on a Euclidean Lattice}}, \href{https://doi.org/10.1103/PhysRevLett.110.262002}{\emph{Phys. Rev. Lett.} {\bfseries 110} (2013) 262002} [\href{https://arxiv.org/abs/1305.1539}{{\ttfamily 1305.1539}}].

\bibitem{Ji:2020ect}
X.~Ji, Y.-S.~Liu, Y.~Liu, J.-H.~Zhang and Y.~Zhao, \emph{{Large-momentum effective theory}}, \href{https://doi.org/10.1103/RevModPhys.93.035005}{\emph{Rev. Mod. Phys.} {\bfseries 93} (2021) 035005} [\href{https://arxiv.org/abs/2004.03543}{{\ttfamily 2004.03543}}].

\bibitem{Radyushkin:2017cyf}
A.V.~Radyushkin, \emph{{Quasi-parton distribution functions, momentum distributions, and pseudo-parton distribution functions}}, \href{https://doi.org/10.1103/PhysRevD.96.034025}{\emph{Phys. Rev. D} {\bfseries 96} (2017) 034025} [\href{https://arxiv.org/abs/1705.01488}{{\ttfamily 1705.01488}}].

\bibitem{Lin:2017snn}
H.-W.~Lin et~al., \emph{{Parton distributions and lattice QCD calculations: a community white paper}}, \href{https://doi.org/10.1016/j.ppnp.2018.01.007}{\emph{Prog. Part. Nucl. Phys.} {\bfseries 100} (2018) 107} [\href{https://arxiv.org/abs/1711.07916}{{\ttfamily 1711.07916}}].

\bibitem{Liu:1993cv}
K.-F.~Liu and S.-J.~Dong, \emph{{Origin of difference between anti-d and anti-u partons in the nucleon}}, \href{https://doi.org/10.1103/PhysRevLett.72.1790}{\emph{Phys. Rev. Lett.} {\bfseries 72} (1994) 1790} [\href{https://arxiv.org/abs/hep-ph/9306299}{{\ttfamily hep-ph/9306299}}].

\bibitem{Detmold:2005gg}
W.~Detmold and C.J.D.~Lin, \emph{{Deep-inelastic scattering and the operator product expansion in lattice QCD}}, \href{https://doi.org/10.1103/PhysRevD.73.014501}{\emph{Phys. Rev. D} {\bfseries 73} (2006) 014501} [\href{https://arxiv.org/abs/hep-lat/0507007}{{\ttfamily hep-lat/0507007}}].

\bibitem{Braun:2007wv}
V.~Braun and D.~M\"uller, \emph{{Exclusive processes in position space and the pion distribution amplitude}}, \href{https://doi.org/10.1140/epjc/s10052-008-0608-4}{\emph{Eur. Phys. J. C} {\bfseries 55} (2008) 349} [\href{https://arxiv.org/abs/0709.1348}{{\ttfamily 0709.1348}}].

\bibitem{Ma:2017pxb}
Y.-Q.~Ma and J.-W.~Qiu, \emph{{Exploring Partonic Structure of Hadrons Using ab initio Lattice QCD Calculations}}, \href{https://doi.org/10.1103/PhysRevLett.120.022003}{\emph{Phys. Rev. Lett.} {\bfseries 120} (2018) 022003} [\href{https://arxiv.org/abs/1709.03018}{{\ttfamily 1709.03018}}].

\bibitem{Chambers:2017dov}
A.J.~Chambers, R.~Horsley, Y.~Nakamura, H.~Perlt, P.E.L.~Rakow, G.~Schierholz et~al., \emph{{Nucleon Structure Functions from Operator Product Expansion on the Lattice}}, \href{https://doi.org/10.1103/PhysRevLett.118.242001}{\emph{Phys. Rev. Lett.} {\bfseries 118} (2017) 242001} [\href{https://arxiv.org/abs/1703.01153}{{\ttfamily 1703.01153}}].

\bibitem{Brodsky:2006uqa}
S.J.~Brodsky and G.F.~de~Teramond, \emph{{Hadronic spectra and light-front wavefunctions in holographic QCD}}, \href{https://doi.org/10.1103/PhysRevLett.96.201601}{\emph{Phys. Rev. Lett.} {\bfseries 96} (2006) 201601} [\href{https://arxiv.org/abs/hep-ph/0602252}{{\ttfamily hep-ph/0602252}}].

\bibitem{1stBLFQ}
J.P.~Vary, H.~Honkanen, J.~Li, P.~Maris, S.J.~Brodsky, A.~Harindranath et~al., \emph{{Hamiltonian light-front field theory in a basis function approach}}, \href{https://doi.org/10.1103/PhysRevC.81.035205}{\emph{Phys. Rev.} {\bfseries C81} (2010) 035205} [\href{https://arxiv.org/abs/0905.1411}{{\ttfamily 0905.1411}}].

\bibitem{Cichy:2018mum}
K.~Cichy and M.~Constantinou, \emph{{A guide to light-cone PDFs from Lattice QCD: an overview of approaches, techniques and results}}, \href{https://doi.org/10.1155/2019/3036904}{\emph{Adv. High Energy Phys.} {\bfseries 2019} (2019) 3036904} [\href{https://arxiv.org/abs/1811.07248}{{\ttfamily 1811.07248}}].

\bibitem{Constantinou:2020pek}
M.~Constantinou, \emph{{The x-dependence of hadronic parton distributions: A review on the progress of lattice QCD}}, \href{https://doi.org/10.1140/epja/s10050-021-00353-7}{\emph{Eur. Phys. J. A} {\bfseries 57} (2021) 77} [\href{https://arxiv.org/abs/2010.02445}{{\ttfamily 2010.02445}}].

\bibitem{Bauer:2022hpo}
C.W.~Bauer et~al., \emph{{Quantum Simulation for High-Energy Physics}}, \href{https://doi.org/10.1103/PRXQuantum.4.027001}{\emph{PRX Quantum} {\bfseries 4} (2023) 027001} [\href{https://arxiv.org/abs/2204.03381}{{\ttfamily 2204.03381}}].

\bibitem{DiMeglio:2023nsa}
A.~Di~Meglio et~al., \emph{{Quantum Computing for High-Energy Physics: State of the Art and Challenges}}, \href{https://doi.org/10.1103/PRXQuantum.5.037001}{\emph{PRX Quantum} {\bfseries 5} (2024) 037001} [\href{https://arxiv.org/abs/2307.03236}{{\ttfamily 2307.03236}}].

\bibitem{Jordan:2011ci}
S.P.~Jordan, K.S.M.~Lee and J.~Preskill, \emph{{Quantum Computation of Scattering in Scalar Quantum Field Theories}}, {\emph{Quant. Inf. Comput.} {\bfseries 14} (2014) 1014} [\href{https://arxiv.org/abs/1112.4833}{{\ttfamily 1112.4833}}].

\bibitem{Jordan:2014tma}
S.P.~Jordan, K.S.M.~Lee and J.~Preskill, \emph{{Quantum Algorithms for Fermionic Quantum Field Theories}},  \href{https://arxiv.org/abs/1404.7115}{{\ttfamily 1404.7115}}.

\bibitem{Jordan:2017lea}
S.P.~Jordan, H.~Krovi, K.S.M.~Lee and J.~Preskill, \emph{{BQP-completeness of Scattering in Scalar Quantum Field Theory}}, \href{https://doi.org/10.22331/q-2018-01-08-44}{\emph{Quantum} {\bfseries 2} (2018) 44} [\href{https://arxiv.org/abs/1703.00454}{{\ttfamily 1703.00454}}].

\bibitem{Lamm:2019uyc}
{\scshape NuQS} collaboration, \emph{{Parton physics on a quantum computer}}, \href{https://doi.org/10.1103/PhysRevResearch.2.013272}{\emph{Phys. Rev. Res.} {\bfseries 2} (2020) 013272} [\href{https://arxiv.org/abs/1908.10439}{{\ttfamily 1908.10439}}].

\bibitem{Perez-Salinas:2020nem}
A.~P\'erez-Salinas, J.~Cruz-Martinez, A.A.~Alhajri and S.~Carrazza, \emph{{Determining the proton content with a quantum computer}}, \href{https://doi.org/10.1103/PhysRevD.103.034027}{\emph{Phys. Rev. D} {\bfseries 103} (2021) 034027} [\href{https://arxiv.org/abs/2011.13934}{{\ttfamily 2011.13934}}].

\bibitem{LiTianyin:2021kcs}
{\scshape QuNu} collaboration, \emph{{Partonic collinear structure by quantum computing}}, \href{https://doi.org/10.1103/PhysRevD.105.L111502}{\emph{Phys. Rev. D} {\bfseries 105} (2022) L111502} [\href{https://arxiv.org/abs/2106.03865}{{\ttfamily 2106.03865}}].

\bibitem{Li:2022lyt}
{\scshape QuNu} collaboration, \emph{{Exploring light-cone distribution amplitudes from quantum computing}}, \href{https://doi.org/10.1007/s11433-023-2120-1}{\emph{Sci. China Phys. Mech. Astron.} {\bfseries 66} (2023) 281011} [\href{https://arxiv.org/abs/2207.13258}{{\ttfamily 2207.13258}}].

\bibitem{LiTianyin:2024nod}
T.~Li, H.~Xing and D.-B.~Zhang, \emph{{Simulating Parton Fragmentation on Quantum Computers}},  \href{https://arxiv.org/abs/2406.05683}{{\ttfamily 2406.05683}}.

\bibitem{Grieninger:2024cdl}
S.~Grieninger, K.~Ikeda and I.~Zahed, \emph{{Quasiparton distributions in massive QED2: Toward quantum computation}}, \href{https://doi.org/10.1103/PhysRevD.110.076008}{\emph{Phys. Rev. D} {\bfseries 110} (2024) 076008} [\href{https://arxiv.org/abs/2404.05112}{{\ttfamily 2404.05112}}].

\bibitem{Grieninger:2024axp}
S.~Grieninger and I.~Zahed, \emph{{Quasi-fragmentation functions in the massive Schwinger model}},  \href{https://arxiv.org/abs/2406.01891}{{\ttfamily 2406.01891}}.

\bibitem{Mueller:2019qqj}
N.~Mueller, A.~Tarasov and R.~Venugopalan, \emph{{Deeply inelastic scattering structure functions on a hybrid quantum computer}}, \href{https://doi.org/10.1103/PhysRevD.102.016007}{\emph{Phys. Rev. D} {\bfseries 102} (2020) 016007} [\href{https://arxiv.org/abs/1908.07051}{{\ttfamily 1908.07051}}].

\bibitem{Echevarria:2020wct}
M.G.~Echevarria, I.L.~Egusquiza, E.~Rico and G.~Schnell, \emph{{Quantum simulation of light-front parton correlators}}, \href{https://doi.org/10.1103/PhysRevD.104.014512}{\emph{Phys. Rev. D} {\bfseries 104} (2021) 014512} [\href{https://arxiv.org/abs/2011.01275}{{\ttfamily 2011.01275}}].

\bibitem{Kreshchuk:2020dla}
M.~Kreshchuk, W.M.~Kirby, G.~Goldstein, H.~Beauchemin and P.J.~Love, \emph{{Quantum simulation of quantum field theory in the light-front formulation}}, \href{https://doi.org/10.1103/PhysRevA.105.032418}{\emph{Phys. Rev. A} {\bfseries 105} (2022) 032418} [\href{https://arxiv.org/abs/2002.04016}{{\ttfamily 2002.04016}}].

\bibitem{Kreshchuk:2020aiq}
M.~Kreshchuk, S.~Jia, W.M.~Kirby, G.~Goldstein, J.P.~Vary and P.J.~Love, \emph{{Simulating Hadronic Physics on NISQ devices using Basis Light-Front Quantization}}, \href{https://doi.org/10.1103/PhysRevA.103.062601}{\emph{Phys. Rev. A} {\bfseries 103} (2021) 062601} [\href{https://arxiv.org/abs/2011.13443}{{\ttfamily 2011.13443}}].

\bibitem{Qian:2021jxp}
W.~Qian, R.~Basili, S.~Pal, G.~Luecke and J.P.~Vary, \emph{{Solving hadron structures using the basis light-front quantization approach on quantum computers}}, \href{https://doi.org/10.1103/PhysRevResearch.4.043193}{\emph{Phys. Rev. Res.} {\bfseries 4} (2022) 043193} [\href{https://arxiv.org/abs/2112.01927}{{\ttfamily 2112.01927}}].

\bibitem{Preskill:2018jim}
J.~Preskill, \emph{{Quantum Computing in the NISQ era and beyond}}, \href{https://doi.org/10.22331/q-2018-08-06-79}{\emph{Quantum} {\bfseries 2} (2018) 79} [\href{https://arxiv.org/abs/1801.00862}{{\ttfamily 1801.00862}}].

\bibitem{Silvi:2017srb}
P.~Silvi, F.~Tschirsich, M.~Gerster, J.~J\"unemann, D.~Jaschke, M.~Rizzi et~al., \emph{{The Tensor Networks Anthology: Simulation techniques for many-body quantum lattice systems}}, \href{https://doi.org/10.21468/SciPostPhysLectNotes.8}{\emph{SciPost Phys. Lect. Notes} {\bfseries 8} (2019) 1} [\href{https://arxiv.org/abs/1710.03733}{{\ttfamily 1710.03733}}].

\bibitem{Banuls:2018jag}
M.C.~Ba\~nuls, K.~Cichy, J.I.~Cirac, K.~Jansen and S.~K\"uhn, \emph{{Tensor Networks and their use for Lattice Gauge Theories}}, \href{https://doi.org/10.22323/1.334.0022}{\emph{PoS} {\bfseries LATTICE2018} (2018) 022} [\href{https://arxiv.org/abs/1810.12838}{{\ttfamily 1810.12838}}].

\bibitem{Orus:2013kga}
R.~Orus, \emph{{A Practical Introduction to Tensor Networks: Matrix Product States and Projected Entangled Pair States}}, \href{https://doi.org/10.1016/j.aop.2014.06.013}{\emph{Annals Phys.} {\bfseries 349} (2014) 117} [\href{https://arxiv.org/abs/1306.2164}{{\ttfamily 1306.2164}}].

\bibitem{mps-reps-gs}
F.~Verstraete and J.I.~Cirac, \emph{{Matrix product states represent ground states faithfully}}, \href{https://doi.org/10.1103/PhysRevB.73.094423}{\emph{Phys. Rev. B} {\bfseries 73} (2006) 094423}.

\bibitem{Barata:2024bzk}
J.a.~Barata, K.~Ikeda, S.~Mukherjee and J.~Raghoonanan, \emph{{Towards Quantum Computing Timelike Hadronic Vacuum Polarization and Light-by-Light Scattering: Schwinger Model Tests}},  \href{https://arxiv.org/abs/2406.03536}{{\ttfamily 2406.03536}}.

\bibitem{Banuls:2024oxa}
M.C.~Ba\~nuls, K.~Cichy, C.J.D.~Lin and M.~Schneider, \emph{{Parton Distribution Functions in the Schwinger Model with Tensor Networks}},  in \emph{{41st International Symposium on Lattice Field Theory}}, 9, 2024 [\href{https://arxiv.org/abs/2409.16996}{{\ttfamily 2409.16996}}].

\bibitem{Banuls:2025wiq}
M.C.~Ban\~{u}ls, K.~Cichy, C.J.D.~Lin and M.~Schneider, \emph{{Parton Distribution Functions in the Schwinger model from Tensor Network States}},  \href{https://arxiv.org/abs/2504.07508}{{\ttfamily 2504.07508}}.

\bibitem{Collins:2011zzd}
J.~Collins, \emph{{Foundations of Perturbative QCD}}, vol.~32 of \emph{Cambridge Monographs on Particle Physics, Nuclear Physics and Cosmology}, Cambridge University Press (7, 2023), \href{https://doi.org/10.1017/9781009401845}{10.1017/9781009401845}.

\bibitem{Lepage:1980fj}
G.P.~Lepage and S.J.~Brodsky, \emph{{Exclusive Processes in Perturbative Quantum Chromodynamics}}, \href{https://doi.org/10.1103/PhysRevD.22.2157}{\emph{Phys. Rev. D} {\bfseries 22} (1980) 2157}.

\bibitem{Efremov:1979qk}
A.V.~Efremov and A.V.~Radyushkin, \emph{{Factorization and Asymptotical Behavior of Pion Form-Factor in QCD}}, \href{https://doi.org/10.1016/0370-2693(80)90869-2}{\emph{Phys. Lett. B} {\bfseries 94} (1980) 245}.

\bibitem{Li:2017mlw}
Y.~Li, P.~Maris and J.P.~Vary, \emph{{Quarkonium as a relativistic bound state on the light front}}, \href{https://doi.org/10.1103/PhysRevD.96.016022}{\emph{Phys. Rev. D} {\bfseries 96} (2017) 016022} [\href{https://arxiv.org/abs/1704.06968}{{\ttfamily 1704.06968}}].

\bibitem{Lan:2019vui}
J.~Lan, C.~Mondal, S.~Jia, X.~Zhao and J.P.~Vary, \emph{{Parton Distribution Functions from a Light Front Hamiltonian and QCD Evolution for Light Mesons}}, \href{https://doi.org/10.1103/PhysRevLett.122.172001}{\emph{Phys. Rev. Lett.} {\bfseries 122} (2019) 172001} [\href{https://arxiv.org/abs/1901.11430}{{\ttfamily 1901.11430}}].

\bibitem{Tang:2019gvn}
S.~Tang, Y.~Li, P.~Maris and J.P.~Vary, \emph{{Heavy-light mesons on the light front}}, \href{https://doi.org/10.1140/epjc/s10052-020-8081-9}{\emph{Eur. Phys. J. C} {\bfseries 80} (2020) 522} [\href{https://arxiv.org/abs/1912.02088}{{\ttfamily 1912.02088}}].

\bibitem{Shuo_Bc}
S.~Tang, Y.~Li, P.~Maris and J.P.~Vary, \emph{{$B_c$ mesons and their properties on the light front}}, \href{https://doi.org/10.1103/PhysRevD.98.114038}{\emph{Phys. Rev.} {\bfseries D98} (2018) 114038} [\href{https://arxiv.org/abs/1810.05971}{{\ttfamily 1810.05971}}].

\bibitem{Qian:2020utg}
W.~Qian, S.~Jia, Y.~Li and J.P.~Vary, \emph{{Light mesons within the basis light-front quantization framework}}, \href{https://doi.org/10.1103/PhysRevC.102.055207}{\emph{Phys. Rev. C} {\bfseries 102} (2020) 055207} [\href{https://arxiv.org/abs/2005.13806}{{\ttfamily 2005.13806}}].

\bibitem{Nambu:1961tp}
Y.~Nambu and G.~Jona-Lasinio, \emph{{Dynamical Model of Elementary Particles Based on an Analogy with Superconductivity. 1.}}, \href{https://doi.org/10.1103/PhysRev.122.345}{\emph{Phys. Rev.} {\bfseries 122} (1961) 345}.

\bibitem{Nambu:1961fr}
Y.~Nambu and G.~Jona-Lasinio, \emph{{Dynamical model of elementary particles based on an analogy with superconductivity. II.}}, \href{https://doi.org/10.1103/PhysRev.124.246}{\emph{Phys. Rev.} {\bfseries 124} (1961) 246}.

\bibitem{Gross1974}
D.J.~Gross and A.~Neveu, \emph{Dynamical symmetry breaking in asymptotically free field theories}, \href{https://doi.org/10.1103/PhysRevD.10.3235}{\emph{Phys. Rev. D} {\bfseries 10} (1974) 3235}.

\bibitem{Shi:2015ufa}
S.~Shi, Y.-C.~Yang, Y.-H.~Xia, Z.-F.~Cui, X.-J.~Liu and H.-S.~Zong, \emph{{Dynamical chiral symmetry breaking in the NJL model with a constant external magnetic field}}, \href{https://doi.org/10.1103/PhysRevD.91.036006}{\emph{Phys. Rev. D} {\bfseries 91} (2015) 036006} [\href{https://arxiv.org/abs/1503.00452}{{\ttfamily 1503.00452}}].

\bibitem{Huang:2001yw}
M.~Huang, P.-f.~Zhuang and W.-q.~Chao, \emph{{Massive quark propagator and competition between chiral and diquark condensate}}, \href{https://doi.org/10.1103/PhysRevD.65.076012}{\emph{Phys. Rev. D} {\bfseries 65} (2002) 076012} [\href{https://arxiv.org/abs/hep-ph/0112124}{{\ttfamily hep-ph/0112124}}].

\bibitem{Thies:2019ejd}
M.~Thies, \emph{{Phase structure of the ( 1+1 )-dimensional Nambu\textendash{}Jona-Lasinio model with isospin}}, \href{https://doi.org/10.1103/PhysRevD.101.014010}{\emph{Phys. Rev. D} {\bfseries 101} (2020) 014010} [\href{https://arxiv.org/abs/1911.11439}{{\ttfamily 1911.11439}}].

\bibitem{Czajka:2021yll}
A.M.~Czajka, Z.-B.~Kang, H.~Ma and F.~Zhao, \emph{{Quantum simulation of chiral phase transitions}}, \href{https://doi.org/10.1007/JHEP08(2022)209}{\emph{JHEP} {\bfseries 08} (2022) 209} [\href{https://arxiv.org/abs/2112.03944}{{\ttfamily 2112.03944}}].

\bibitem{Borsanyi:2010cj}
S.~Borsanyi, G.~Endrodi, Z.~Fodor, A.~Jakovac, S.D.~Katz, S.~Krieg et~al., \emph{{The QCD equation of state with dynamical quarks}}, \href{https://doi.org/10.1007/JHEP11(2010)077}{\emph{JHEP} {\bfseries 11} (2010) 077} [\href{https://arxiv.org/abs/1007.2580}{{\ttfamily 1007.2580}}].

\bibitem{Chakrabarti:2003wi}
D.~Chakrabarti, A.~Harindranath and J.P.~Vary, \emph{{A Study of q anti-q states in transverse lattice QCD using alternative fermion formulations}}, \href{https://doi.org/10.1103/PhysRevD.69.034502}{\emph{Phys. Rev. D} {\bfseries 69} (2004) 034502} [\href{https://arxiv.org/abs/hep-ph/0309317}{{\ttfamily hep-ph/0309317}}].

\bibitem{Jordan:1928wi}
P.~Jordan and E.P.~Wigner, \emph{{About the Pauli exclusion principle}}, \href{https://doi.org/10.1007/BF01331938}{\emph{Z. Phys.} {\bfseries 47} (1928) 631}.

\bibitem{Zohar:2012ts}
E.~Zohar, J.I.~Cirac and B.~Reznik, \emph{{Simulating (2+1)-Dimensional Lattice QED with Dynamical Matter Using Ultracold Atoms}}, \href{https://doi.org/10.1103/PhysRevLett.110.055302}{\emph{Phys. Rev. Lett.} {\bfseries 110} (2013) 055302} [\href{https://arxiv.org/abs/1208.4299}{{\ttfamily 1208.4299}}].

\bibitem{Brennen:2015pgn}
G.K.~Brennen, G.~Pupillo, E.~Rico, T.M.~Stace and D.~Vodola, \emph{{Loops and Strings in a Superconducting Lattice Gauge Simulator}}, \href{https://doi.org/10.1103/PhysRevLett.117.240504}{\emph{Phys. Rev. Lett.} {\bfseries 117} (2016) 240504} [\href{https://arxiv.org/abs/1512.06565}{{\ttfamily 1512.06565}}].

\bibitem{Lamm:2018siq}
H.~Lamm and S.~Lawrence, \emph{{Simulation of Nonequilibrium Dynamics on a Quantum Computer}}, \href{https://doi.org/10.1103/PhysRevLett.121.170501}{\emph{Phys. Rev. Lett.} {\bfseries 121} (2018) 170501} [\href{https://arxiv.org/abs/1806.06649}{{\ttfamily 1806.06649}}].

\bibitem{Zohar:2019tsa}
E.~Zohar, \emph{{Local Manipulation and Measurement of Nonlocal Many-Body Operators in Lattice Gauge Theory Quantum Simulators}}, \href{https://doi.org/10.1103/PhysRevD.101.034518}{\emph{Phys. Rev. D} {\bfseries 101} (2020) 034518} [\href{https://arxiv.org/abs/1911.11156}{{\ttfamily 1911.11156}}].

\bibitem{nielsen_chuang_2010}
M.A.~Nielsen and I.L.~Chuang, \emph{Quantum Computation and Quantum Information: 10th Anniversary Edition}, Cambridge University Press (2010), \href{https://doi.org/10.1017/CBO9780511976667}{10.1017/CBO9780511976667}.

\bibitem{wiersema2020exploring}
R.~Wiersema, C.~Zhou, Y.~de~Sereville, J.F.~Carrasquilla, Y.B.~Kim and H.~Yuen, \emph{Exploring entanglement and optimization within the hamiltonian variational ansatz}, {\emph{PRX quantum} {\bfseries 1} (2020) 020319}.

\bibitem{Choi:2020pdg}
K.~Choi, D.~Lee, J.~Bonitati, Z.~Qian and J.~Watkins, \emph{{Rodeo Algorithm for Quantum Computing}}, \href{https://doi.org/10.1103/PhysRevLett.127.040505}{\emph{Phys. Rev. Lett.} {\bfseries 127} (2021) 040505} [\href{https://arxiv.org/abs/2009.04092}{{\ttfamily 2009.04092}}].

\bibitem{Chakraborty:2020uhf}
B.~Chakraborty, M.~Honda, T.~Izubuchi, Y.~Kikuchi and A.~Tomiya, \emph{{Classically emulated digital quantum simulation of the Schwinger model with a topological term via adiabatic state preparation}}, \href{https://doi.org/10.1103/PhysRevD.105.094503}{\emph{Phys. Rev. D} {\bfseries 105} (2022) 094503} [\href{https://arxiv.org/abs/2001.00485}{{\ttfamily 2001.00485}}].

\bibitem{Lin:2020zni}
L.~Lin and Y.~Tong, \emph{{Near-optimal ground state preparation}}, \href{https://doi.org/10.22331/q-2020-12-14-372}{\emph{Quantum} {\bfseries 4} (2020) 372} [\href{https://arxiv.org/abs/2002.12508}{{\ttfamily 2002.12508}}].

\bibitem{Dong:2022mmq}
Y.~Dong, L.~Lin and Y.~Tong, \emph{{Ground-State Preparation and Energy Estimation on Early Fault-Tolerant Quantum Computers via Quantum Eigenvalue Transformation of Unitary Matrices}}, \href{https://doi.org/10.1103/PRXQuantum.3.040305}{\emph{PRX Quantum} {\bfseries 3} (2022) 040305} [\href{https://arxiv.org/abs/2204.05955}{{\ttfamily 2204.05955}}].

\bibitem{Kane:2023jdo}
C.F.~Kane, N.~Gomes and M.~Kreshchuk, \emph{{Nearly optimal state preparation for quantum simulations of lattice gauge theories}}, \href{https://doi.org/10.1103/PhysRevA.110.012455}{\emph{Phys. Rev. A} {\bfseries 110} (2024) 012455} [\href{https://arxiv.org/abs/2310.13757}{{\ttfamily 2310.13757}}].

\bibitem{Hastings:2007iok}
M.B.~Hastings, \emph{{An area law for one-dimensional quantum systems}}, \href{https://doi.org/10.1088/1742-5468/2007/08/P08024}{\emph{J. Stat. Mech.} {\bfseries 0708} (2007) P08024} [\href{https://arxiv.org/abs/0705.2024}{{\ttfamily 0705.2024}}].

\bibitem{vidal-erg}
G.~Vidal, \emph{{Entanglement Renormalization: an introduction}},  \href{https://arxiv.org/abs/0912.1651}{{\ttfamily 0912.1651}}.

\bibitem{mcclean2017hybrid}
J.R.~McClean, M.E.~Kimchi-Schwartz, J.~Carter and W.A.~De~Jong, \emph{Hybrid quantum-classical hierarchy for mitigation of decoherence and determination of excited states}, {\emph{Physical Review A} {\bfseries 95} (2017) 042308}.

\bibitem{White:1992zz}
S.R.~White, \emph{{Density matrix formulation for quantum renormalization groups}}, \href{https://doi.org/10.1103/PhysRevLett.69.2863}{\emph{Phys. Rev. Lett.} {\bfseries 69} (1992) 2863}.

\bibitem{White:1993zza}
S.R.~White, \emph{{Density-matrix algorithms for quantum renormalization groups}}, \href{https://doi.org/10.1103/PhysRevB.48.10345}{\emph{Phys. Rev. B} {\bfseries 48} (1993) 10345}.

\bibitem{Schollwock2005}
U.~Schollw\"ock, \emph{The density-matrix renormalization group}, \href{https://doi.org/10.1103/RevModPhys.77.259}{\emph{Rev. Mod. Phys.} {\bfseries 77} (2005) 259}.

\bibitem{schollwock2011density}
U.~Schollw{\"o}ck, \emph{The density-matrix renormalization group in the age of matrix product states}, {\emph{Annals of physics} {\bfseries 326} (2011) 96}.

\bibitem{frank-dmrg}
F.~Verstraete, T.~Nishino, U.~Schollw{\"o}ck, M.C.~Ba{\~n}uls, G.K.-L.~Chan and M.E.~Stoudenmire, \emph{Density matrix renormalization group, 30 years on}, {\emph{Nature Reviews Physics} {\bfseries 5} (2023) 273}.

\bibitem{ITensor:2022}
M.~Fishman, S.R.~White and E.M.~Stoudenmire, \emph{The itensor software library for tensor network calculations}, \href{https://doi.org/10.21468/SciPostPhysCodeb.4}{\emph{SciPost Phys. Codebases} (2022) 4}.

\bibitem{ostlund1995thermodynamic}
S.~{\"O}stlund and S.~Rommer, \emph{Thermodynamic limit of density matrix renormalization}, {\emph{Physical review letters} {\bfseries 75} (1995) 3537}.

\bibitem{dmrg-on-tpul}
M.~Ganahl, J.~Beall, M.~Hauru, A.G.M.~Lewis, T.~Wojno, J.H.~Yoo et~al., \emph{{Density Matrix Renormalization Group with Tensor Processing Units}}, \href{https://doi.org/10.1103/PRXQuantum.4.010317}{\emph{PRX Quantum} {\bfseries 4} (2023) 010317} [\href{https://arxiv.org/abs/2204.05693}{{\ttfamily 2204.05693}}].

\bibitem{Vidal:2003lvx}
G.~Vidal, \emph{{Efficient simulation of one-dimensional quantum many-body systems}}, \href{https://doi.org/10.1103/PhysRevLett.93.040502}{\emph{Phys. Rev. Lett.} {\bfseries 93} (2004) 040502} [\href{https://arxiv.org/abs/quant-ph/0310089}{{\ttfamily quant-ph/0310089}}].

\bibitem{verstraete2004matrix}
F.~Verstraete, J.J.~Garcia-Ripoll and J.I.~Cirac, \emph{Matrix product density operators: Simulation of finite-temperature and dissipative systems}, {\emph{Physical review letters} {\bfseries 93} (2004) 207204}.

\bibitem{Haegeman:2011zz}
J.~Haegeman, J.I.~Cirac, T.J.~Osborne, I.~Pizorn, H.~Verschelde and F.~Verstraete, \emph{{Time-Dependent Variational Principle for Quantum Lattices}}, \href{https://doi.org/10.1103/PhysRevLett.107.070601}{\emph{Phys. Rev. Lett.} {\bfseries 107} (2011) 070601} [\href{https://arxiv.org/abs/1103.0936}{{\ttfamily 1103.0936}}].

\bibitem{haegeman2016unifying}
J.~Haegeman, C.~Lubich, I.~Oseledets, B.~Vandereycken and F.~Verstraete, \emph{Unifying time evolution and optimization with matrix product states}, {\emph{Physical Review B} {\bfseries 94} (2016) 165116}.

\bibitem{tdvp-scaling}
C.~Hubig, J.~Haegeman and U.~Schollw\"ock, \emph{Error estimates for extrapolations with matrix-product states}, \href{https://doi.org/10.1103/PhysRevB.97.045125}{\emph{Phys. Rev. B} {\bfseries 97} (2018) 045125}.

\bibitem{vidal-tebd-many-body}
G.~Vidal, \emph{{Efficient simulation of one-dimensional quantum many-body systems}}, \href{https://doi.org/10.1103/PhysRevLett.93.040502}{\emph{Phys. Rev. Lett.} {\bfseries 93} (2004) 040502} [\href{https://arxiv.org/abs/quant-ph/0310089}{{\ttfamily quant-ph/0310089}}].

\bibitem{vidal-tebd-qc}
G.~Vidal, \emph{{Efficient Classical Simulation of Slightly Entangled Quantum Computations}}, \href{https://doi.org/10.1103/PhysRevLett.91.147902}{\emph{Phys. Rev. Lett.} {\bfseries 91} (2003) 147902}.

\bibitem{Kim:2013etb}
H.~Kim and D.A.~Huse, \emph{{Ballistic Spreading of Entanglement in a Diffusive Nonintegrable System}}, \href{https://doi.org/10.1103/PhysRevLett.111.127205}{\emph{Phys. Rev. Lett.} {\bfseries 111} (2013) 127205}.

\bibitem{Schuch_2008}
N.~Schuch, M.M.~Wolf, K.G.H.~Vollbrecht and J.I.~Cirac, \emph{On entropy growth and the hardness of simulating time evolution}, \href{https://doi.org/10.1088/1367-2630/10/3/033032}{\emph{New Journal of Physics} {\bfseries 10} (2008) 033032}.

\bibitem{Calabrese:2005in}
P.~Calabrese and J.L.~Cardy, \emph{{Evolution of entanglement entropy in one-dimensional systems}}, \href{https://doi.org/10.1088/1742-5468/2005/04/P04010}{\emph{J. Stat. Mech.} {\bfseries 0504} (2005) P04010} [\href{https://arxiv.org/abs/cond-mat/0503393}{{\ttfamily cond-mat/0503393}}].

\bibitem{Schachenmayer:2013}
J.~Schachenmayer, B.P.~Lanyon, C.F.~Roos and A.J.~Daley, \emph{Entanglement growth in quench dynamics with variable range interactions}, \href{https://doi.org/10.1103/PhysRevX.3.031015}{\emph{Phys. Rev. X} {\bfseries 3} (2013) 031015}.

\bibitem{itensor-tdvp}
M.~Yang and S.R.~White, \emph{Time-dependent variational principle with ancillary krylov subspace}, \href{https://doi.org/10.1103/PhysRevB.102.094315}{\emph{Phys. Rev. B} {\bfseries 102} (2020) 094315}.

\bibitem{Ikeda:2024rzv}
K.~Ikeda, Z.-B.~Kang, D.E.~Kharzeev, W.~Qian and F.~Zhao, \emph{{Real-time chiral dynamics at finite temperature from quantum simulation}}, \href{https://doi.org/10.1007/JHEP10(2024)031}{\emph{JHEP} {\bfseries 10} (2024) 031} [\href{https://arxiv.org/abs/2407.21496}{{\ttfamily 2407.21496}}].

\bibitem{Trotter1959}
H.F.~Trotter, \emph{On the product of semi-groups of operators}, {\emph{Proceedings of the American Mathematical Society} {\bfseries 10} (1959) 545}.

\bibitem{childs2019nearly}
A.M.~Childs and Y.~Su, \emph{Nearly optimal lattice simulation by product formulas}, {\emph{Physical review letters} {\bfseries 123} (2019) 050503}.

\bibitem{childs2021theory}
A.M.~Childs, Y.~Su, M.C.~Tran, N.~Wiebe and S.~Zhu, \emph{Theory of trotter error with commutator scaling}, {\emph{Physical Review X} {\bfseries 11} (2021) 011020}.

\bibitem{Qiskit}
M.S.~Anis et~al., \emph{{Qiskit: An Open-source Framework for Quantum Computing}},  {2021}.
\newblock {10.5281/zenodo.2573505}.

\bibitem{E791:2000xcx}
{\scshape E791} collaboration, \emph{{Direct measurement of the pion valence quark momentum distribution, the pion light cone wave function squared}}, \href{https://doi.org/10.1103/PhysRevLett.86.4768}{\emph{Phys. Rev. Lett.} {\bfseries 86} (2001) 4768} [\href{https://arxiv.org/abs/hep-ex/0010043}{{\ttfamily hep-ex/0010043}}].

\bibitem{Conway:1989fs}
J.S.~Conway et~al., \emph{{Experimental Study of Muon Pairs Produced by 252-GeV Pions on Tungsten}}, \href{https://doi.org/10.1103/PhysRevD.39.92}{\emph{Phys. Rev. D} {\bfseries 39} (1989) 92}.

\bibitem{Jia:2018qee}
Y.~Jia, S.~Liang, X.~Xiong and R.~Yu, \emph{{Partonic quasidistributions in two-dimensional QCD}}, \href{https://doi.org/10.1103/PhysRevD.98.054011}{\emph{Phys. Rev. D} {\bfseries 98} (2018) 054011} [\href{https://arxiv.org/abs/1804.04644}{{\ttfamily 1804.04644}}].

\bibitem{Brodsky:1997de}
S.J.~Brodsky, H.-C.~Pauli and S.S.~Pinsky, \emph{{Quantum chromodynamics and other field theories on the light cone}}, \href{https://doi.org/10.1016/S0370-1573(97)00089-6}{\emph{Phys. Rept.} {\bfseries 301} (1998) 299} [\href{https://arxiv.org/abs/hep-ph/9705477}{{\ttfamily hep-ph/9705477}}].

\bibitem{Brodsky:2014yha}
S.J.~Brodsky, G.F.~de~Teramond, H.G.~Dosch and J.~Erlich, \emph{{Light-Front Holographic QCD and Emerging Confinement}}, \href{https://doi.org/10.1016/j.physrep.2015.05.001}{\emph{Phys. Rept.} {\bfseries 584} (2015) 1} [\href{https://arxiv.org/abs/1407.8131}{{\ttfamily 1407.8131}}].

\bibitem{Vega:2009zb}
A.~Vega, I.~Schmidt, T.~Branz, T.~Gutsche and V.E.~Lyubovitskij, \emph{{Meson wave function from holographic models}}, \href{https://doi.org/10.1103/PhysRevD.80.055014}{\emph{Phys. Rev. D} {\bfseries 80} (2009) 055014} [\href{https://arxiv.org/abs/0906.1220}{{\ttfamily 0906.1220}}].

\bibitem{Swarnkar:2015osa}
R.~Swarnkar and D.~Chakrabarti, \emph{{Meson structure in light-front holographic QCD}}, \href{https://doi.org/10.1103/PhysRevD.92.074023}{\emph{Phys. Rev. D} {\bfseries 92} (2015) 074023} [\href{https://arxiv.org/abs/1507.01568}{{\ttfamily 1507.01568}}].

\bibitem{Jia:2018ary}
S.~Jia and J.P.~Vary, \emph{{Basis light front quantization for the charged light mesons with color singlet Nambu–Jona-Lasinio interactions}}, \href{https://doi.org/10.1103/PhysRevC.99.035206}{\emph{Phys. Rev.} {\bfseries C99} (2019) 035206} [\href{https://arxiv.org/abs/1811.08512}{{\ttfamily 1811.08512}}].

\bibitem{Lan:2019img}
J.~Lan, C.~Mondal, M.~Li, Y.~Li, S.~Tang, X.~Zhao et~al., \emph{{Parton Distribution Functions of Heavy Mesons on the Light Front}}, \href{https://doi.org/10.1103/PhysRevD.102.014020}{\emph{Phys. Rev. D} {\bfseries 102} (2020) 014020} [\href{https://arxiv.org/abs/1911.11676}{{\ttfamily 1911.11676}}].

\bibitem{Verstraete:2004cf}
F.~Verstraete and J.I.~Cirac, \emph{{Renormalization algorithms for quantum-many body systems in two and higher dimensions}},  \href{https://arxiv.org/abs/cond-mat/0407066}{{\ttfamily cond-mat/0407066}}.

\bibitem{verstraete2004valence}
F.~Verstraete and J.I.~Cirac, \emph{Valence-bond states for quantum computation}, {\emph{Physical Review A—Atomic, Molecular, and Optical Physics} {\bfseries 70} (2004) 060302}.

\bibitem{Shi:2005ctp}
Y.Y.~Shi, L.M.~Duan and G.~Vidal, \emph{{Classical simulation of quantum many-body systems with a tree tensor network}}, \href{https://doi.org/10.1103/PhysRevA.74.022320}{\emph{Phys. Rev. A} {\bfseries 74} (2006) 022320} [\href{https://arxiv.org/abs/quant-ph/0511070}{{\ttfamily quant-ph/0511070}}].

\bibitem{Magnifico:2020bqt}
G.~Magnifico, T.~Felser, P.~Silvi and S.~Montangero, \emph{{Lattice quantum electrodynamics in (3+1)-dimensions at finite density with tensor networks}}, \href{https://doi.org/10.1038/s41467-021-23646-3}{\emph{Nature Commun.} {\bfseries 12} (2021) 3600} [\href{https://arxiv.org/abs/2011.10658}{{\ttfamily 2011.10658}}].

\bibitem{LevinNave:2007}
M.~Levin and C.P.~Nave, \emph{{Tensor renormalization group approach to 2D classical lattice models}}, \href{https://doi.org/10.1103/PhysRevLett.99.120601}{\emph{Phys. Rev. Lett.} {\bfseries 99} (2007) 120601} [\href{https://arxiv.org/abs/cond-mat/0611687}{{\ttfamily cond-mat/0611687}}].

\bibitem{tnr}
G.~Evenbly and G.~Vidal, \emph{{Tensor Network Renormalization}}, \href{https://doi.org/10.1103/PhysRevLett.115.180405}{\emph{Phys. Rev. Lett.} {\bfseries 115} (2015) 180405}.

\end{thebibliography}\endgroup
\end{document}